\documentclass[twocolumn]{aastex631} 

\usepackage{amsmath}
\usepackage{bm}
\usepackage{CJK}
\usepackage{graphbox}

\graphicspath{{./}{fig/}}

\newcommand{\mat}[1]{\bm{#1}}
\renewcommand{\vec}[1]{\bm{#1}}

\journalinfo{The Astronomical Journal, in press} 
\shorttitle{Preview of C/2021~A1 and Venus Encounter}
\shortauthors{Zhang et al.}

\received{2021 June 20}
\revised{2021 July 20}
\accepted{2021 July 24}

\renewcommand{\edit}[2]{#2} 

\begin{document}
\begin{CJK*}{UTF8}{gbsn}

\title{Preview of Comet C/2021~A1 (Leonard) and Its Encounter with Venus}

\author[0000-0002-6702-191X]{Qicheng Zhang}
\affiliation{Division of Geological and Planetary Sciences, California Institute of Technology, Pasadena, CA 91125, USA}

\author[0000-0002-4838-7676]{Quanzhi Ye (叶泉志)}
\affiliation{Department of Astronomy, University of Maryland, College Park, MD 20742, USA}

\author[0000-0003-2527-1475]{Shreyas Vissapragada}
\affiliation{Division of Geological and Planetary Sciences, California Institute of Technology, Pasadena, CA 91125, USA}

\author[0000-0003-2781-6897]{Matthew M. Knight}
\affiliation{Department of Physics, United States Naval Academy, Annapolis, MD 21402, USA}

\author[0000-0002-4767-9861]{Tony L. Farnham}
\affiliation{Department of Astronomy, University of Maryland, College Park, MD 20742, USA}

\correspondingauthor{Qicheng Zhang}
\email{qicheng@cometary.org}

\begin{abstract}
Long period comet C/2021~A1 (Leonard) will approach Venus to within 0.029~au on 2021~December~18 \edit1{and may subsequently graze the planet with its dust trail} less than two days later. We observed C/2021~A1 with the Lowell Discovery Telescope on 2021~January~13 and March~3, as well as with the Palomar Hale Telescope on 2021~March~20, while the comet was inbound at heliocentric distances of $r=4.97$~au, 4.46~au, and 4.28~au, respectively. Tail morphology suggests that the dust is optically dominated by $\sim$0.1--1~mm radius grains produced in the prior year. Neither narrowband imaging photometry nor spectrophotometry reveal any definitive gas emission, placing $3\sigma$ upper bounds on CN production of ${\lesssim}10^{23}$~molec~s$^{-1}$ at both of the latter two epochs. Trajectory analysis indicates that large ($\gtrsim$1~mm) grains ejected at extremely large heliocentric distances ($r\gtrsim30$~au) are most strongly favored to reach Venus. The flux of such meteors on Venus, and thus their potential direct or indirect observability, is highly uncertain as the comet's dust production history is poorly constrained at these distances, but will likely fall well below \edit2{the meteor flux} from comet C/2013~A1 (Siding Spring)'s closer encounter to Mars in 2014, and thus poses negligible risk to \edit1{any} spacecraft \edit1{in orbit around Venus}. Dust produced in previous apparitions will not likely contribute substantially to the meteor flux, nor will dust from any future activity apart from an unlikely high speed ($\gtrsim$0.5~km~s$^{-1}$) dust outburst prior to the comet reaching $r\approx2$~au in 2021~September.
\end{abstract}


\section{Introduction}

\object{C/2021~A1 (Leonard)} is a long period comet discovered by G.~J. Leonard on 2021~January~3 as part of the Catalina Sky Survey \citep{leonard2021a}. Prediscovery observations were subsequently located to extend the astrometric observation arc back to 2020~April~11 and indicated that it would approach to within 0.029~au of Venus on 2021~December~18 \citep{leonard2021b}, a distance closer than that of any reliably constrained encounter of a long period comet to Earth.\footnote{\url{https://www.minorplanetcenter.net/iau/lists/ClosestComets.html}} More remarkably, Venus will \edit1{pass only $\sim$50,000~km---the \emph{minimum orbit intersection distance} (MOID)---from the comet's heliocentric orbit on 2021~December~19 at a point only three days behind the nucleus}. This passage is notable because the principal action of solar radiation pressure on dust grains ejected from a comet nucleus is to confine them to a size-sorted dust fan, with the smallest grains (radius $a_d\sim1$~$\mu$m) rapidly dispersed in the anti-sunward direction and the largest grains ($a_d\gtrsim1$~mm) more slowly concentrated into a narrow trail along the comet's heliocentric orbit, lagging behind the nucleus \citep{finson1968}. The exceptionally close approach of Venus to a portion of the comet's orbit close behind the nucleus therefore raises the possibility that large dust grains from C/2021~A1 may reach Venus to produce a meteor shower.

Meteors are more commonly associated with short period comets, which rapidly fill their physically \edit1{smaller orbits} over many revolutions with dust trails that may be directly observed \citep[e.g.,][]{sykes1992} or otherwise inferred from meteor outbursts when Earth crosses their orbits \citep[e.g.,][]{kresak1993}. Past trails of longer period comets are more sparsely distributed and thinly stretched over their larger orbits, although meteors from such trails are also occasionally detected \citep[e.g.,][]{lyytinen2003}. \citet{beech1998}, \citet{christou2010}, and others have additionally investigated such periodic meteoroid streams for their potential to generate annual meteor showers on Venus, although such meteor showers have yet to be clearly observed.

Although not a dynamically new comet, C/2021~A1 follows a very large orbit with inbound barycentric semimajor axis $a_\mathrm{in}\approx1900$~au, corresponding to a dynamical lifetime on the order of one orbital period \citep{krolikowska2017}, which precludes the comet from having developed a \edit1{discernible}, periodic meteoroid stream \edit1{\citep{jenniskens2021}}. Rather, we are interested in the potential for meteors produced in the ongoing apparition---while the dust is still concentrated behind the nucleus---which is a far more rare occurrence that requires not only a close approach of a comet but one immediately followed by the planet crossing the comet's orbit behind the nucleus. A nearly as close 0.031~au encounter of C/1983~H1 (IRAS--Araki--Alcock) to Earth---the closest known of a long period comet to Earth---thus produced minimal enhancement in meteor activity, as Earth crossed ahead of rather than behind the nucleus\edit1{, missing the dust trail} \citep{ohtsuka1991}. \edit1{Likewise, the 0.07~au encounter of C/1976~E1 (Bradfield) to Venus---the previous closest by a long period comet to Venus reported by JPL SBDB\footnote{\url{https://ssd-api.jpl.nasa.gov/doc/cad.html}}---likely produced no meteor activity as Venus passed far interior to the comet's orbit, where no recent dust from the comet could plausibly reach.}

Nonetheless, the favorability of C/2021~A1's encounter with Venus is not unprecedented. Most notably, comet C/2013~A1 (Siding Spring) passed 140,000~km from Mars on 2014~October~19 \citep{farnocchia2016}. Following this encounter, Mars crossed the dust trail (MOID $\sim30{,}000$~km) at a point less than three hours behind the nucleus, intercepting a sizeable fraction of the large, 1--10~mm grains released from the nucleus earlier at a heliocentric distance of $r\sim22.5$~au \citep{farnocchia2014}. Ultraviolet spectroscopy by the MAVEN spacecraft in orbit around Mars at the time subsequently revealed the appearance of a temporary metallic vapor layer in the atmosphere with a density consistent with ${\sim}10^4$~kg of deposited dust \citep{schneider2015}, while associated perturbations to the ionosphere were independently detected by multiple other spacecraft and instruments \citep{benna2015,restano2015,sanchezcano2020}.

In the following sections, we present early imagery and spectroscopy of C/2021~A1 to provide an initial characterization of the comet and enable a preliminary comparison with other comets. We then discuss the orbital setup, and both the dynamical and physical requirements for dust to reach Venus, and compare these results with those of the prior encounter of C/2013~A1 with Mars. We also briefly speculate on the prospects for detecting meteors on Venus either by direct observation or indirectly through the presence of a meteoritic layer or associated ionospheric perturbations as seen on Mars.

\section{Observations}

\begin{deluxetable*}{lcccccccc}
\tablecaption{Observations of C/2021~A1\label{tab:obs}}

\tablecolumns{9}
\tablehead{
\colhead{Time} & \colhead{Instrument} & \colhead{Band} & \colhead{Aperture} & \colhead{Flux} & \colhead{$Af\rho$} & \colhead{$A(0^\circ)f\rho$} & \colhead{Gas Flux} & \colhead{Gas Production}\\
\colhead{(UT)} & \colhead{} & \colhead{} & \colhead{} & \colhead{(mag)\tablenotemark{a}} & \colhead{(m)} & \colhead{(m)\tablenotemark{b}} & \colhead{(W~m$^{-2}$)\tablenotemark{c}} & \colhead{(molec~s$^{-1}$)}
}

\startdata
\noalign{\vspace{0.8ex}}
\multicolumn{9}{c}{2021~Jan~13 ($r=4.97$~au\tablenotemark{d}, $\varDelta=4.71$~au\tablenotemark{e}, $\alpha=11^\circ\llap{.}2$\tablenotemark{f})}\\
\noalign{\vspace{0.8ex}}
\hline
13:01--13:16 & LDT/LMI & \textit{VR}\tablenotemark{g} & $2''$ radius & $19.4\pm0.1$ & $2.2\pm0.2$ & $3.3\pm0.3$ & \nodata & \nodata\\
& & & $5''$ radius & $18.9\pm0.1$ & $1.3\pm0.1$ & $2.0\pm0.2$ & \nodata & \nodata\\
\hline
\noalign{\vspace{0.8ex}}
\multicolumn{9}{c}{2021~Mar~3 ($r=4.46$~au\tablenotemark{d}, $\varDelta=3.86$~au\tablenotemark{e}, $\alpha=10^\circ\llap{.}9$\tablenotemark{f})}\\
\noalign{\vspace{0.8ex}}
\hline
08:13--08:43 & LDT/LMI & SDSS $r'$ & $2''$ radius & $19.03\pm0.08$ & $2.0\pm0.2$ & $3.0\pm0.3$ & \nodata & \nodata\\
& & HB BC & & $19.90\pm0.06$ & $1.7\pm0.1$ & $2.6\pm0.2$ & \nodata  & \nodata\\
& & HB RC & & $18.43\pm0.05$ & $1.9\pm0.1$ & $2.9\pm0.1$ & \nodata  & \nodata\\
& & SDSS $r'$ & $5''$ radius & $18.46\pm0.05$ & $1.35\pm0.07$ & $2.0\pm0.1$ & \nodata & \nodata\\
& & HB BC & & $19.26\pm0.05$ & $1.23\pm0.06$ & $1.84\pm0.09$ & \nodata  & \nodata\\
& & HB RC & & $17.74\pm0.04$ & $1.46\pm0.06$ & $2.18\pm0.09$ & \nodata  & \nodata\\
& & CN ($\Delta v=0$)\tablenotemark{h} & & \nodata & \nodata & \nodata & ${<}10^{-18.4}$ & ${<}10^{23.0}$\\
\hline
\noalign{\vspace{0.8ex}}
\multicolumn{9}{c}{2021~Mar~20 ($r=4.28$~au\tablenotemark{d}, $\varDelta=3.70$~au\tablenotemark{e}, $\alpha=11^\circ\llap{.}7$\tablenotemark{f})}\\
\noalign{\vspace{0.8ex}}
\hline
04:44--05:25 & P200/WIRC & 2MASS $J$ & $2''$ radius & $17.42\pm0.06$ & $3.0\pm0.2$ & $4.6\pm0.3$ & \nodata & \nodata\\
& & & $5''$ radius & $16.88\pm0.08$ & $2.0\pm0.2$ & $3.1\pm0.3$ & \nodata & \nodata\\
& & & $10''\times10''$ & $16.86\pm0.09$ & $1.8\pm0.2$ & $2.8\pm0.3$ & \nodata & \nodata\\
\hline
07:52--08:14 & P200/DBSP\tablenotemark{i} & Bessell $B$ & $10''\times10''$ & $19.19\pm0.01$ & $1.13\pm0.01$ & $1.73\pm0.01$ & \nodata & \nodata\\
& & Bessell $V$ & & $18.42\pm0.01$ & $1.29\pm0.01$ & $1.99\pm0.02$ & \nodata & \nodata\\
& & Bessell $R$ & & $17.98\pm0.01$ & $1.35\pm0.01$ & $2.08\pm0.02$ & \nodata & \nodata\\
& & Bessell $I$ & & $17.59\pm0.01$ & $1.42\pm0.01$ & $2.18\pm0.02$ & \nodata & \nodata\\
& & SDSS $g'$ & & $18.80\pm0.01$ & $1.18\pm0.01$ & $1.81\pm0.02$ & \nodata & \nodata\\
& & SDSS $r'$ & & $18.19\pm0.01$ & $1.33\pm0.01$ & $2.05\pm0.02$ & \nodata & \nodata\\
& & SDSS $i'$ & & $18.03\pm0.01$ & $1.38\pm0.01$ & $2.13\pm0.02$ & \nodata & \nodata\\
& & HB UC & & $20.33\pm0.07$ & $0.97\pm0.07$ & $1.48\pm0.10$ & \nodata & \nodata\\
& & HB BC & & $19.02\pm0.02$ & $1.14\pm0.02$ & $1.75\pm0.03$ & \nodata & \nodata\\
& & HB GC & & $18.39\pm0.02$ & $1.29\pm0.02$ & $1.99\pm0.04$ & \nodata & \nodata\\
& & HB RC & & $17.53\pm0.01$ & $1.37\pm0.01$ & $2.10\pm0.02$ & \nodata & \nodata\\
& & CN ($\Delta v=0$) & & $20.13\pm0.05$ & \nodata & \nodata & ${<}10^{-18.1}$ & ${<}10^{23.3}$\\
& & C$_3$ (Swings) & & $19.59\pm0.03$ & \nodata & \nodata & ${<}10^{-17.4}$ & ${<}10^{22.4}$\\
& & C$_2$ ($\Delta v=0$) & & $18.50\pm0.02$ & \nodata & \nodata & ${<}10^{-17.8}$ & ${<}10^{23.0}$\\
\enddata
\tablenotetext{a}{Native magnitude for each standard photometric system (SDSS: AB, 2MASS: Vega = 0, Bessell: Vega = 0.03, HB: HD 191263 = 6.19); HB magnitudes before continuum subtraction for gas emission bands.}
\tablenotetext{b}{Converted from observed $Af\rho$ with the Schleicher--Marcus phase function \citep{schleicher2011}.}
\tablenotetext{c}{Full band continuum subtracted gas flux. Upper limits are $3\sigma$.}
\tablenotetext{d}{Heliocentric distance.}
\tablenotetext{e}{Geocentric distance.}
\tablenotetext{f}{Phase angle.}
\tablenotetext{g}{Calibrated as SDSS $r'$.}
\tablenotetext{h}{Derived from aperture differencing (see \S~\ref{sec:gas}), and converted to $5''$ radius equivalent aperture.}
\tablenotetext{i}{Relative uncertainties given for DBSP colors. Absolute flux and $Af\rho$ for stated aperture estimated to be $\pm$5\%.}
\end{deluxetable*}

\begin{figure*}
\centering
\includegraphics[width=0.32\linewidth]{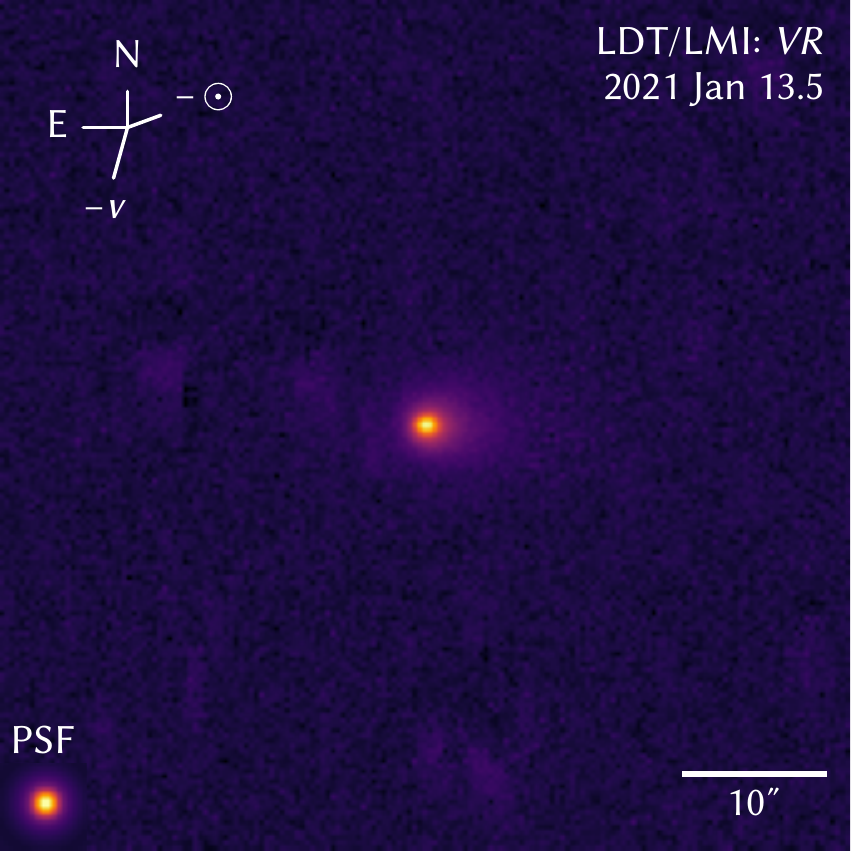}
\includegraphics[width=0.32\linewidth]{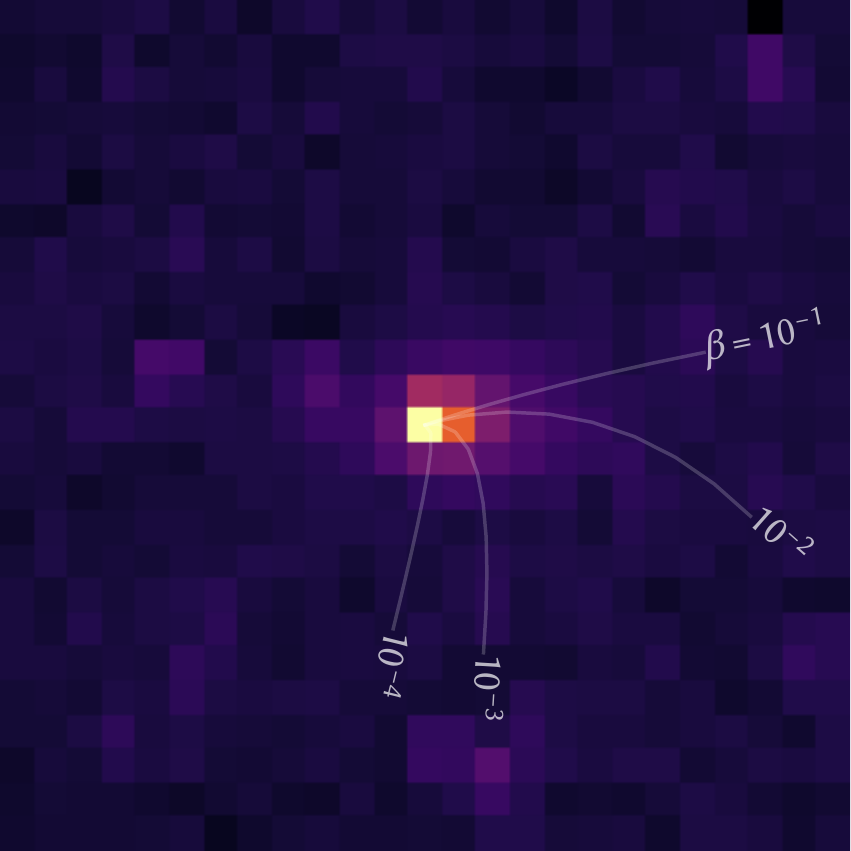}
\includegraphics[width=0.32\linewidth]{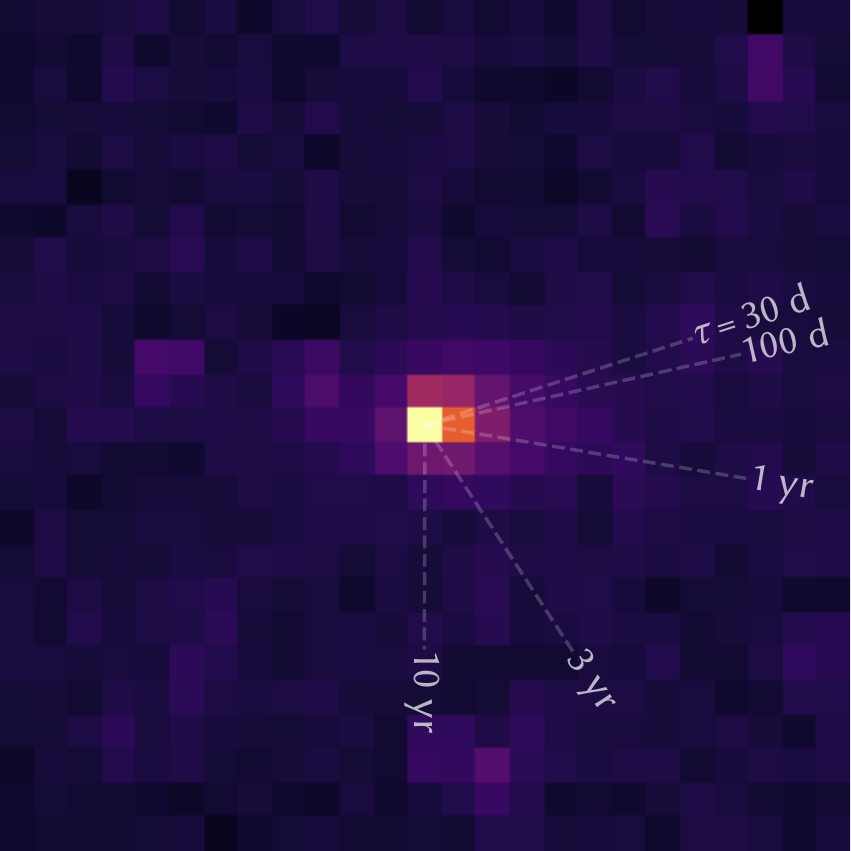}
\includegraphics[width=0.32\linewidth]{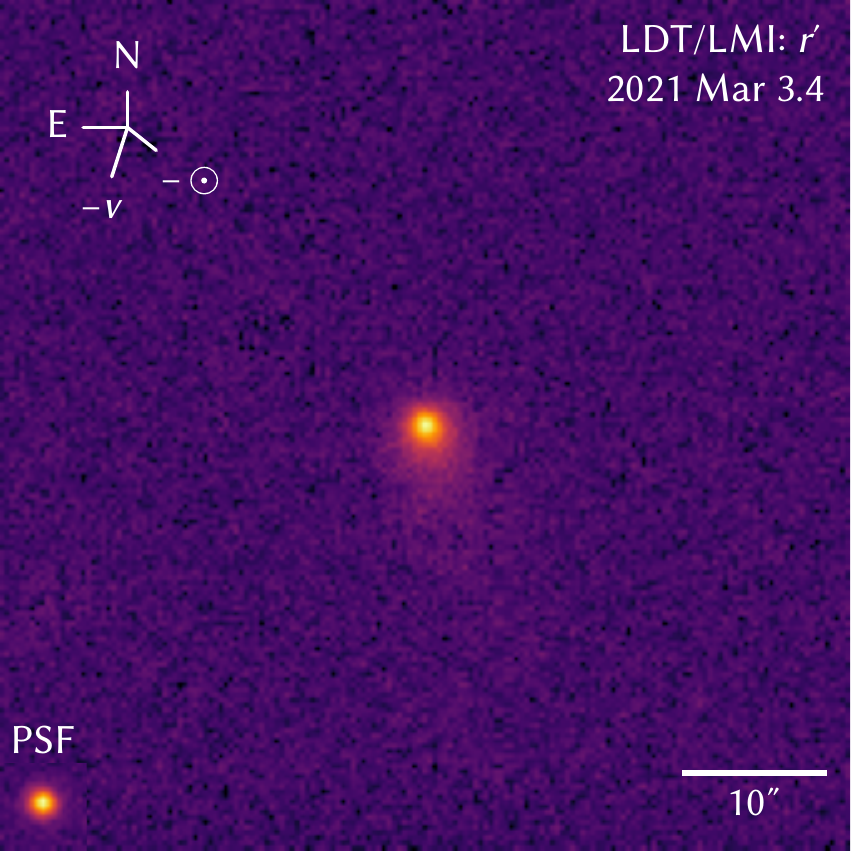}
\includegraphics[width=0.32\linewidth]{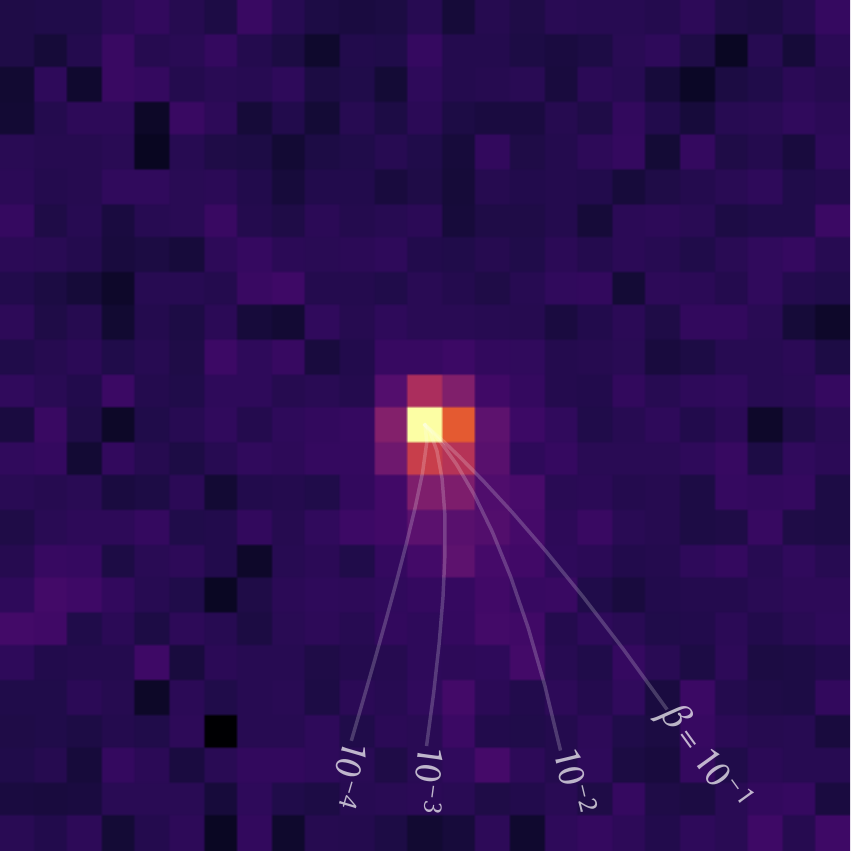}
\includegraphics[width=0.32\linewidth]{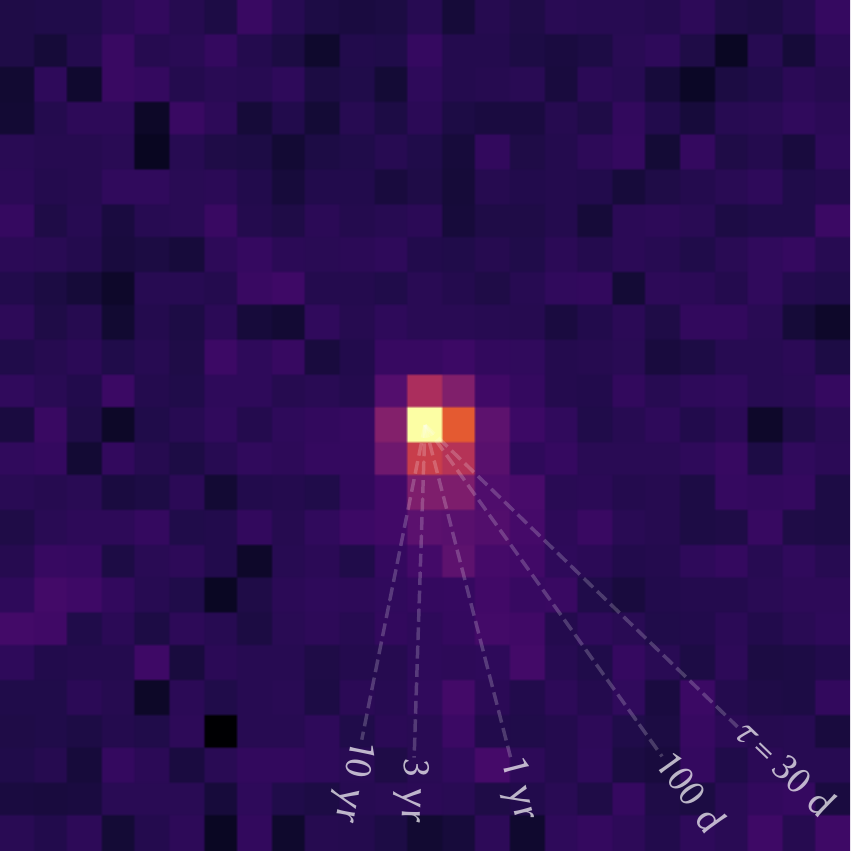}
\includegraphics[width=0.32\linewidth]{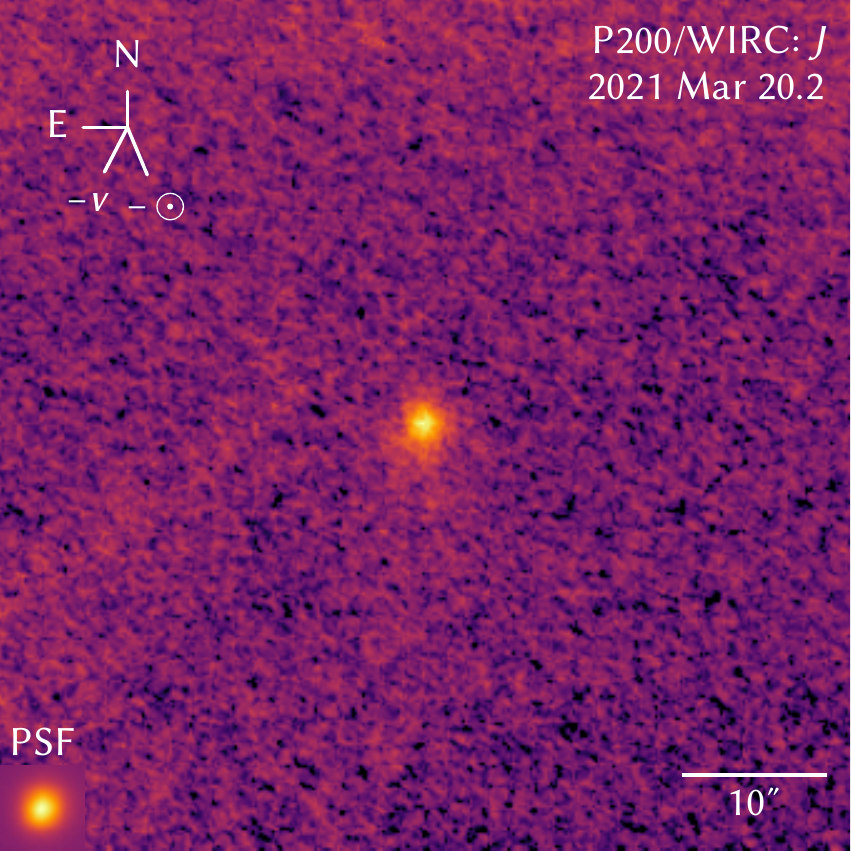}
\includegraphics[width=0.32\linewidth]{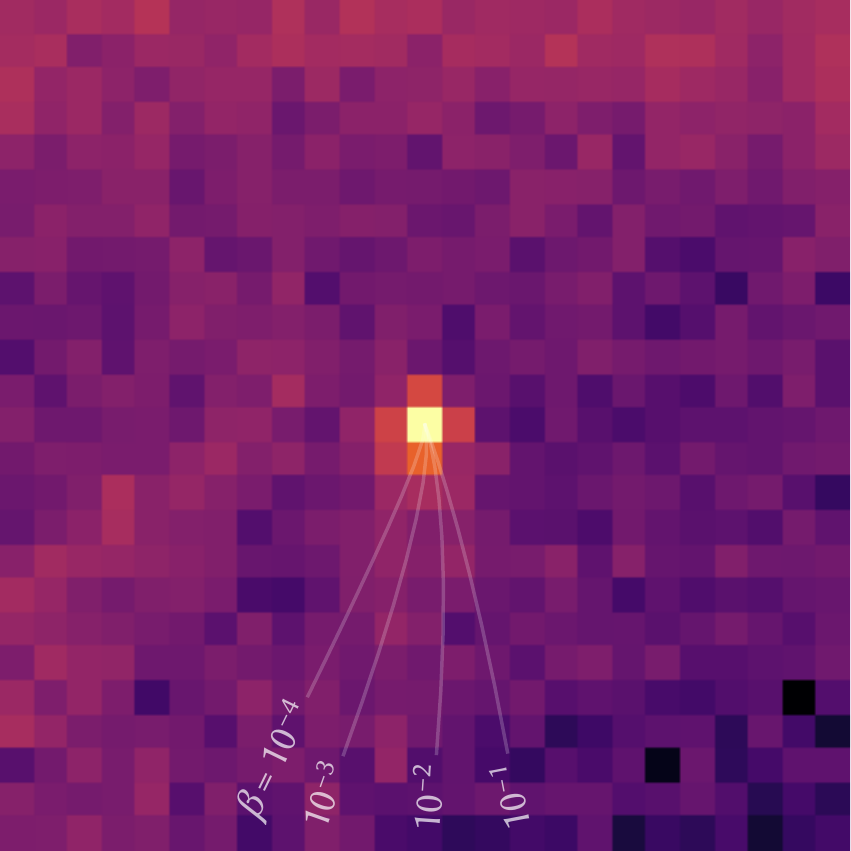}
\includegraphics[width=0.32\linewidth]{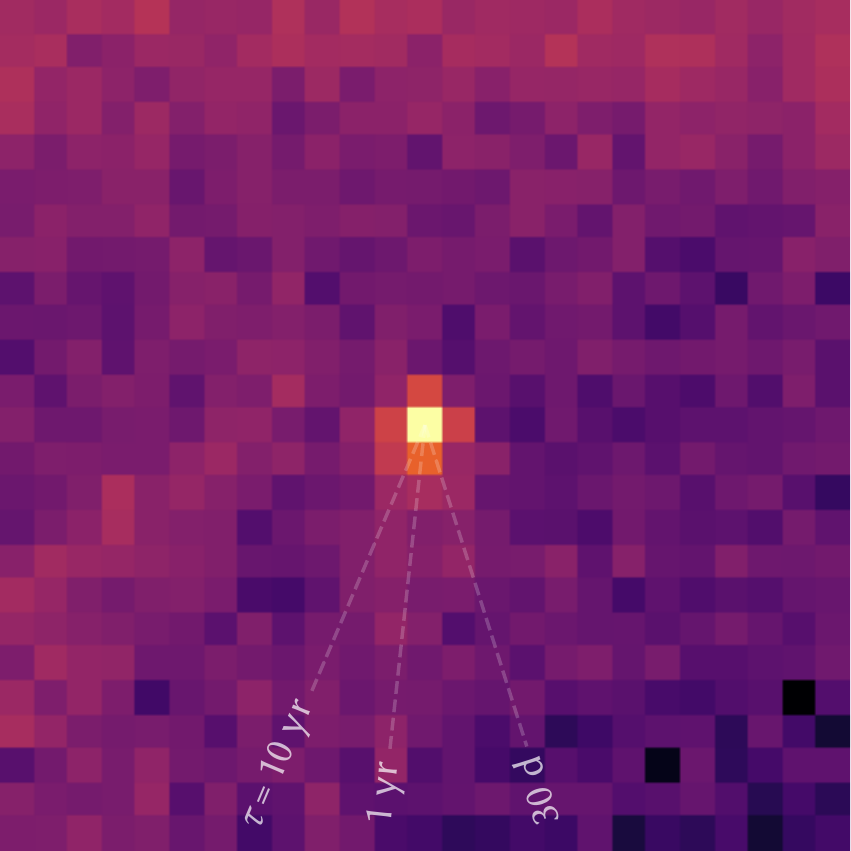}
\caption{Stacked images from LDT/LMI through the \textit{VR} filter on 2021~January~13 (top row), through $r'$ on 2021~March~3 (middle row) and from P200/WIRC through $J$ on 2021~March~20 (bottom row). The left panels show the original image with a measured PSF scaled to the brightness of the comet for comparison, and with $-\odot$ and $-v$ indicating the anti-sunward and negative heliocentric velocity directions, respectively. The middle and right panels show the same images binned to 2.4~arcsec~px$^{-1}$ to emphasize the tail. Syndynes (middle) and synchrones (right) are overlaid on both frames, and suggest that the brightness of the tail is predominantly from dust produced within the previous year with $\beta\sim10^{-2}$--$10^{-3}$, roughly corresponding to $a_d\sim0.1$--1~mm. Note that the $\beta=1$ syndyne (not plotted) is tightly compressed against the $\beta=0.1$ syndyne in all epochs, so we cannot exclude a small brightness contribution from $\beta>0.1$ grains.}
\label{fig:tail}
\end{figure*}

We observed C/2021~A1 with the Lowell Discovery Telescope (LDT) on 2021~January~13 and 2021~March~3 for optical broadband and narrowband imaging, and with the Palomar Hale Telescope (P200) on 2021~March~20 for near-infrared ($J$) imaging and optical spectroscopy. These observations and the photometry derived from them are summarized in Table~\ref{tab:obs}, with associated images of the comet at all epochs shown in Figure~\ref{fig:tail}.

\subsection{Lowell Discovery Telescope (LDT)}
\label{sec:ldt}

We observed the comet using the Large Monolithic Imager \citep[LMI;][]{massey2013} on LDT. The LMI has a field of view of $12'\llap{.}3\times12'\llap{.}3$ and a pixel scale of $0''\llap{.}36$ after an on-chip $3\times3$ binning. For the 2021~January~13 run, we collected three usable images through the \textit{VR} ($V+R$; introduced by \citealt{jewitt1996}) filter only, each with 300~s exposure. For the 2021~March~3 run, we collected two sets of images: broadband imagery in $r'$, as well as narrowband imagery through the CN, BC (blue continuum), and RC (red continuum) filters of the HB set \citep{farnham2000}. The center wavelength of these filters are 387, 445, and 713 nm, respectively, selected to contain the violet CN ($\Delta v=0$) emission bands and nearly gas-free continuum points. We began and ended the observation with one 30~s $r'$ image, and cycled through RC, CN and BC three times, with 180~s exposure for each filter each time. This strategy counters brightness variations in time, including from changing airmass and from any passing cirrus, although we note that the sky was clear during the observation. We also dithered the pointing between each exposure in order to mitigate detector defects such as dead pixels and pixel-to-pixel variation.

We then performed bias subtraction and flat-field correction on the images collected from both nights. Pixels below the 5$^\text{th}$ and above the 95$^\text{th}$ percentiles, excluding a region of $5''$ in radius centered on the comet ($10''$ for the \textit{VR} images from 2021~January~13 due to the images being deeper), were masked to remove background stars. Frames were then median-combined into the final stacked images \edit1{for all} filters.

The \textit{VR} and $r'$ images were photometrically calibrated using the Pan-STARRS1 DR1 catalog \citep{chambers2016}, with \textit{VR} approximated as $r'$. The narrowband images were photometrically calibrated using the HB magnitudes of \edit1{eight} field stars computed from their spectra observed by LAMOST \citep{cui2012}. These HB magnitudes were derived by convolving the transmission profile of each HB filter with the LAMOST spectrum of the star calibrated using the Pan-STARRS1 DR1 catalog, since the LAMOST catalog does not provide absolutely calibrated fluxes. To exclude variable stars and other poor quality spectra, we only used stars with broadband color error---estimated as the standard deviation of the zero-point magnitudes of the spectrum measured for $g'$, $r'$, and $i'$---below 0.1~mag. We also collected images of HB standards during our observation, but encountered unexpected instrumental behavior that prevented us from using these images to perform absolute calibration.

\subsection{Palomar Hale Telescope (P200)}

We used the Wide-field Infrared Camera \citep[WIRC;][]{wilson2003} on P200 to collect $J$-band imagery of the comet. We took $46\times60$~s exposures spread over a nine-point ($3\times3$) square dither pattern, with a grid spacing of $90''$. We performed background subtraction for frames at each dither point by using the median of all frames at the other eight dither positions as the background frame. We then applied flat-field correction with dome flats taken at the beginning of the night, then astrometrically and photometrically solved each frame with field stars from the Gaia EDR3 \citep{gaia2021} and 2MASS \citep{skrutskie2006} catalogs, respectively, before aligning and stacking all frames on the position of the comet.

We \edit1{also} used the Double Spectrograph \citep[DBSP;][]{oke1982} to collect low resolution spectroscopy of the comet. We used a standard configuration capturing wavelengths of $\sim$300--550~nm on the blue side and $\sim$550--1100~nm on the red side, as well as a wide $10''$ slit in order to maximize the signal from the comet and improve sensitivity to any diffuse gas emission\edit1{, and took $4\times300$~s exposures through both channels}. The slit was oriented in the east-west direction, roughly perpendicular to the tail. As the slit is much wider than the typical atmospheric dispersion over the covered wavelength range, dispersion slit losses are negligible irrespective of slit orientation. We additionally observed the nearby ($12^\circ$ away) spectrophotometric standard star Grw+70~5824 (airmass 1.35) \edit1{and took $4\times60$~s exposures} with the same configuration immediately prior to observing the comet (airmass 1.14) for flux and sky calibration.

\begin{figure*}
\centering
\includegraphics[align=c,width=0.55\linewidth]{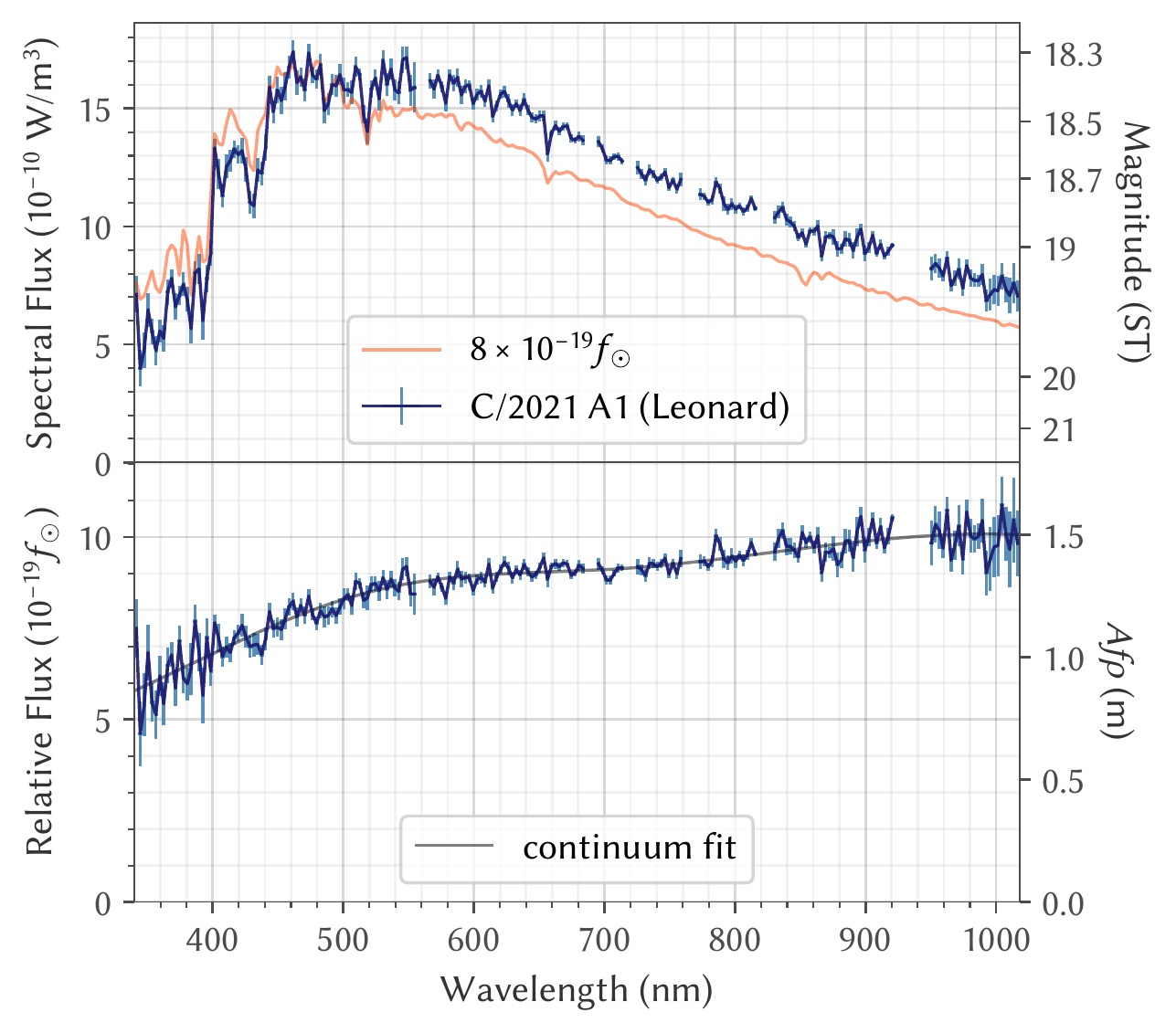}
\hspace{0.05\linewidth}
\includegraphics[align=c,width=0.32\linewidth]{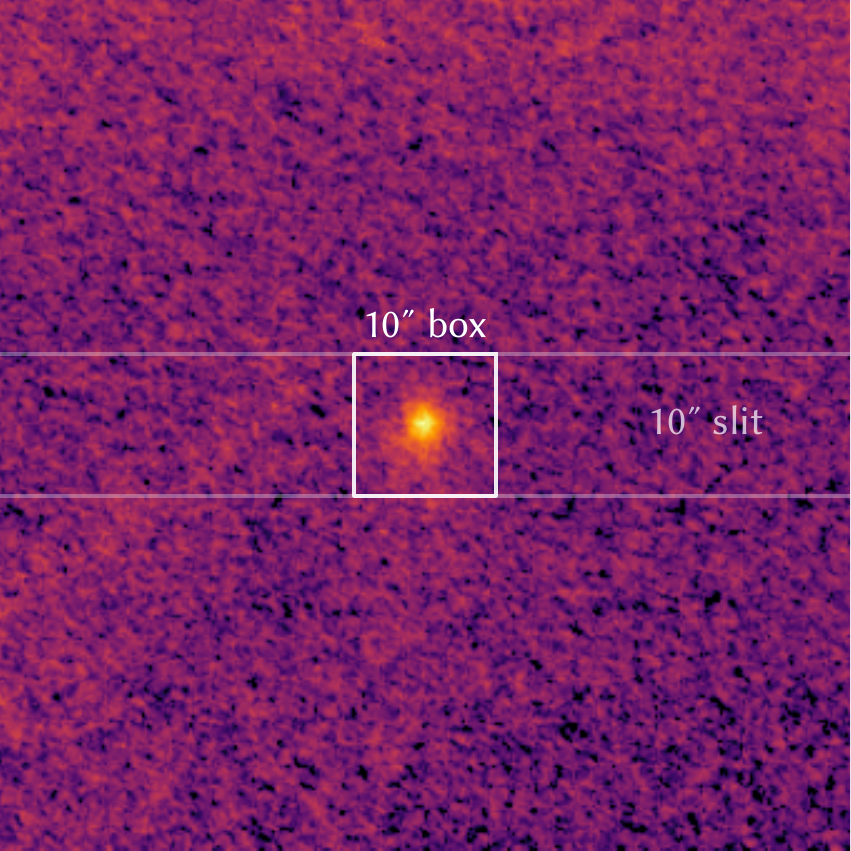}
\caption{\emph{Left:} Spectrum of C/2021~A1 by P200/DBSP as absolute flux per wavelength interval (top) and relative to the solar spectral flux (bottom), the latter a measure of relative reflectance. The best fit continuum, used for gas production constraints, is also shown for reference. \emph{Right:} Illustration showing the relative size and orientation of the $10''$ slit and extraction box aperture overlaid on the P200/WIRC image from Figure~\ref{fig:tail}.}
\label{fig:spec}
\end{figure*}

We used PypeIt \citep{prochaska2020} to extract a $10''$ wide region centered on the comet for spectrophotometry, corresponding to a $10''\times10''$ box aperture. Figure~\ref{fig:spec} shows the \edit1{summed and} fully calibrated spectrum, binned to 3~nm resolution, as well as an overlay of the slit and extraction aperture over the WIRC image. The solar spectrum of \citet{meftah2018} is provided as comparison and was divided into the absolute comet spectrum to produce a \edit1{relative} reflectance spectrum. Flux calibration produced a number of artifacts near major telluric absorption features on the red side, likely due to the difference in airmass and brightness profile between the calibration star and comet, and the most prominent of these artifacts have been masked out \edit1{to aid determination of dust color}. A detector artifact on the red camera near 550~nm was also masked out. We used PypeIt's standard background subtraction routine which measures the sky outside the $10''$ extraction box of our C/2021~A1 spectrum, except for the wavelength range 380--520~nm where we instead subtracted a scaled version of the measured Grw+70~5824 sky background in order to avoid subtraction of any spatially extended gas emission in this wavelength range. We do not consider any gas emission features beyond this range, and the contemporaneous imagery shows the dust coma to be clearly contained within the $10''$ box, so no features of interest should be included in the subtracted background. Dust colors and gas production limits are discussed further in \S~\ref{sec:dust} and \S~\ref{sec:gas}, respectively.

\subsection{Dust Properties}
\label{sec:dust}

Under the formalism of \citet{finson1968}, the spatial position of a dust grain released from the nucleus at zero velocity ($v_d=0$~m~s$^{-1}$) is uniquely determined by a grid of two properties: (1) the time since the grain was released, $\tau$, and (2) the size parameter $\beta$, defined as the ratio of force of solar radiation pressure acting on the grain to that of solar gravitation. For typical low albedo grains of a bulk density $\rho_d\sim0.5$~g~cm$^{-3}$, the effective grain radius $a_d$ is approximately related to $\beta$ by the inverse relation $a_d\sim1~\mu\mathrm{m}/\beta$. We note that estimates of the mean $\rho_d$ vary substantially, with values reported in the range of $\sim$0.1--4~g~cm$^{-3}$ \citep{lasue2009} with individual comets often producing a mixture of both compact and fluffy grains \citep{kolokolova2018}. Differences in $\rho_d$ do not affect our results, except in this rough conversion between $\beta$ and $a_d$.

Both stacked images in Figure~\ref{fig:tail} are accompanied by binned and stretched versions that emphasize the low surface brightness tail, which are overlaid with a selection of \emph{syndynes} and \emph{synchrones}---curves where $v_d=0$ dust grains of equal $\beta$ and equal $\tau$ fall, respectively. The actual non-zero $v_d$ means the dust with these properties will only approximately fall along these curves, and may be biased to one side by anisotropic dust production, but comparison of tail morphology with respect to the grid of $\beta$ and $\tau$ still enables a useful if rudimentary initial characterization of the dust properties in the tail well away from the coma.

The tail within the observed ${\sim}30''$ of the coma is cleanly confined to $\beta>10^{-4}$, with the surface brightness maximum near $\beta\sim10^{-2}$--$10^{-3}$ at $\tau\lesssim1$~yr, corresponding to $a_d\sim0.1$--1~mm. This dust is much larger than typically observed from comets nearer to the Sun ($r\sim1$~au) which are usually brightest in the $\beta\sim0.1$--1 range \citep{fulle2004}. However, it is comparable with the optically dominant $\beta\sim10^{-2}$ dust grains observed from other comets at similar $r$, including C/2013~A1 \citep{ye2014}.

We note that the dust that is optically most prominent in the tail is not necessarily representative of the mass distribution of the dust at any point in time. In particular, while much of the observed $\beta\sim10^{-2}$--$10^{-3}$ dust in the tail was produced at $\tau\sim1$~yr ($r\sim8$~au), the larger $\beta\sim10^{-4}$ dust produced at the same time would still have yet to move out of the coma, and the lower brightness of the tail over this syndyne at similar distances from the coma may merely reflect a lower overall dust production rate at $\tau\gtrsim10$~yr ($r\gtrsim30$~au) than at $\sim$1~yr, rather than an absence of larger grains being produced. Detailed dust simulations fitted to the surface brightness of the tail are needed to probe these grains, but are beyond the scope of the present analysis.

The spectrum in Figure~\ref{fig:spec} shows no obvious gas emission lines, so measured colors represent those of the dust. The dust is redder at shorter wavelengths, with a spectral slope of $16\%\pm3\%$ per 100~nm between the HB BC and GC bands (445--526~nm) dropping to $3\%\pm1\%$ per 100~nm between GC and RC (526--713~nm). The $B-V=0.76\pm0.01$, $V-R=0.44\pm0.01$, and $R-I=0.39\pm0.01$ derived from the spectrum are nearly indistinguishable from the mean colors reported by \citet{jewitt2015} for a sample of distant long period comets ($0.78\pm0.02$, $0.47\pm0.02$, and $0.42\pm0.03$, respectively). The derived F438W--F606W (438--606~nm) slope of $10\%\pm1\%$ per 100~nm is also substantially redder than the $5.0\%\pm0.3\%$ per 100~nm found by \citet{li2014} for C/2013~A1---a dynamically new comet, unlike C/2021~A1---at similar $r\sim4$~au.

\citet{ahearn1984} introduced the use of the $Af\rho$ parameter as a proxy for the dust production of a comet that requires minimal assumptions about the physical makeup of the dust to compute, unlike the true dust production rate. The $Af\rho$ measured inside a photometric aperture centered on the nucleus is the product of the dust's apparent albedo, $A$, a filling factor, $f$, representing the ratio of scattering cross section area to aperture area, and the radius, $\rho$, of the aperture, and is minimally dependent on $\rho$ for comets with steady state dust production where the dust brightness approximately follows a $1/\rho$ profile.

We measured the $Af\rho$ directly with circular aperture photometry for images from all three epochs. Additionally, although intended for circular apertures, we also estimated the $Af\rho$ over the range of our spectrum in Figure~\ref{fig:spec} and Table~\ref{tab:obs} by treating the $10''\times10''$ box aperture as equivalent to a $\rho=5''\llap{.}7$ circular aperture, as both would capture an identical flux from a $1/\rho$ brightness distribution. We measured $Af\rho$ in $\mathrm{\textit{VR}}\sim r'$ ($\sim$625~nm) of $1.3\pm0.1$~m on 2021~January~13, and in HB RC ($\sim$713~nm) of $1.46\pm0.06$~m on 2021~March~3 and $1.37\pm0.05$~m on 2021~March~20 within effective $\rho\approx5''\approx15{,}000$~km. Since $A$ is strongly dependent on phase angle $\alpha$, the $Af\rho$ should normally be corrected to the same $\alpha$ before comparison.

We corrected our $Af\rho$ to $\alpha=0^\circ$ using the Schleicher--Marcus phase function \citep{schleicher2011}, as an approximation for the true wavelength dependent phase function of C/2021~A1, and \edit1{found} corresponding $A(0^\circ)f\rho$ of $2.0\pm0.2$~m, $2.2\pm0.1$~m and $2.1\pm0.1$~m in $r'$ within $\rho\approx5''$---indistinguishable from constant over our observing period. The $Af\rho$ within a smaller $\rho=2''\approx6{,}000$~km is $\sim$50--70\% larger in images at all epochs, and likely better reflects the dust production for the extremely small coma that is ${<}5''$ in radius for comparison with other comets, but are noisier and more strongly affected by seeing conditions. These $A(0^\circ)f\rho$ values are all several times smaller than the $\sim$15~m measured for C/2013~A1 at similar $r$ \citep{ye2014}, which---given the fairly similar $\beta$ of dust observed in the tail---suggests that the dust production of C/2021~A1 is likely lower than that of C/2013~A1 at this distance, although quantitative constraints on the relative dust albedo and size distributions is required to make a definitive conclusion.

\subsection{Gas Production}
\label{sec:gas}

We did not robustly detect any gas emission at either epoch. We used the simple single component model of \citet{haser1957} to place approximate upper limits on the production of CN, C$_3$, and C$_2$ with a generous assumed outflow speed of 1~km~s$^{-1}$, and adopted the efficiency factors and scale lengths compiled by \citet{ahearn1995}, except for the formation length of CN. \citet{fray2005} notes that breakdown of an unknown parent---possibly dust---is responsible for the short CN formation lengths of ${\sim}10^{4}$~km typically observed and assumed for comets near $r\sim1$~au, while HCN photodissociation---which has a much longer expected scale length of ${\sim}8\times10^4~\mathrm{km}\times(r/1~\mathrm{au})^2$ \citep{hanni2020}---should be the principal source of CN at the $r>3$~au of our observations, and we adopt this latter value as the formation length of CN in our model. We note that the final production rate estimates increase with increasing formation scale and outflow speed, so an overestimated formation scale and outflow speed provides conservative (i.e., higher) upper limits on gas production.

Since the CN formation scale is larger than the LDT/LMI field-of-view, we adopt the aperture differencing technique described in \citet{ye2021} to derive the CN production rate from the LDT images. This technique compares the measured difference between two apertures against the modeled value and therefore, relaxes the need to observe a separate background field. After experimenting with apertures of different radii, we used two annuli with inner/outer radii of $10''/15''$ and $45''/60''$ to provide the most stringent constraint. The inner annulus purposely excludes the dust coma to avoid needing a continuum correction. Such a correction would require extrapolating the RC and BC fluxes to the CN bandpass, which is inaccurate if continuum color slope is not constant over the full CN--RC range. We also note that ${\sim}60''$ is the largest usable aperture due to uncorrected stray light patterns starting ${\sim}90''$ from the nucleus that reach ${\sim}1\%$ of the background level. Such patterns were not noted in \citet{ye2021}, likely due to their much shorter exposures and much brighter target. We determined the $3\sigma$ upper limit of the excess flux per wavelength interval within the inner annulus above the surface brightness of the outer annulus to be $8.0\times10^{-11}~\mathrm{W~m^{-3}}$, corresponding to a $3\sigma$ CN production limit of $Q(\mathrm{CN})<10^{23.0}~\mathrm{molec~s^{-1}}$.

We also derived upper limits for CN, C$_3$, and C$_2$ production on 2021~March~20 from the DBSP spectrum. We first used emcee \citep{foreman-mackey2013} to fit a five-point cubic spline function to our reflectance spectrum. To avoid including the gas emission of interest into this fit, we fitted only points in and near the HB continuum bands, and points longward of RC. We approximated the flux uncertainty of all bins as normally distributed and independent, used uniform priors, and verified burn-in completion by visual inspection of the parameter chains. We then measured the residual flux---accounting for both the flux and continuum uncertainty---through the HB CN, C$_3$, and C$_2$ and bandpasses, scaled to the full band CN ($\Delta v=0$), C$_3$ (Swings), and C$_2$ ($\Delta v=0$) fluxes following \citet{farnham2000}, then converted the $3\sigma$ upper bounds to column densities and subsequently production rates by the simple \citet{haser1957} models described earlier. This procedure yielded $3\sigma$ upper bounds of $Q(\mathrm{CN})<10^{23.3}$~molec~s$^{-1}$, $Q(\mathrm{C}_3)<10^{22.4}$~molec~s$^{-1}$, and $Q(\mathrm{C}_2)<10^{23.0}$~molec~s$^{-1}$, which are roughly equally stringent for typical cometary abundances \citep{ahearn1995}.

While no gas emission was detected, we can compare the $Q$(CN) upper limits with the $A(0^\circ)f\rho=2.9\pm0.1$~m measured by LDT in RC within $\rho=2''$, and define a ratio CN/$Af\rho\equiv Q(\mathrm{CN})/(A(0^\circ)f\rho)$ for which we find CN/$Af\rho<10^{22.5}$~molec~s$^{-1}$~m$^{-1}$. This bound is several times smaller than the smallest CN/$Af\rho$ measured by \citet{ahearn1995} after an approximate correction of the latter to $0^\circ$ phase angle, although the vast majority of measurements were made at much smaller $r$. \citet{knight2014} reported a detection of CN/$Af\rho=10^{24.1\pm0.2}$~molec~s$^{-1}$~m$^{-1}$ for the dynamically new sungrazing comet C/2012~S1 (ISON) at $r=4.6$~au pre-perihelion. Similarly, \citet{rauer1997} found CN/$Af\rho=10^{23.5\pm0.5}$~molec~s$^{-1}$~m$^{-1}$ for the very large C/1995~O1 (Hale--Bopp) at $r=4.6$~au. No measurements of $Q$(CN) were made of C/2013~A1 until it had already reached $r=2.4$~au, where \citet{opitom2016} found CN/$Af\rho=10^{24.0\pm0.1}$~molec~s$^{-1}$~m$^{-1}$.

The interpretation and implications of the low CN/$Af\rho$ observed for C/2021~A1 are unclear without a bigger sample of measurements for comparable long period comets observed at similar $r$. The absence of the non-HCN parent of CN at $r>3$~au is likely at least partially responsible. We also speculate that HCN---which sublimates at a lower temperature than H$_2$O---may be somewhat depleted by thermal fractionation from the outer layers of H$_2$O ice on the comet, the latter of which may already be weakly sublimating and driving dust production at $r\sim5$~au \citep{meech2004}. Another possibility is that our $Af\rho$ apertures are sampling an abundance of old, slow-moving grains in the coma and in the portion of the tail behind the coma, thus leading the measured $Af\rho$ to be more reflective of the comet's accumulated dust production history rather than its ongoing dust production. In all of these cases, we anticipate that CN/$Af\rho$ will rise to more typical values as the non-HCN source begins to contribute, more efficient H$_2$O sublimation clears any fractionated layer, old dust grains leave the coma, and changing viewing geometry moves the tail out from behind the coma as the comet approaches the Sun.

\section{Dust Dynamics}

The $b$-plane of a close encounter of a minor to a major planetary body---in our case, C/2021~A1 to Venus---is defined as the plane containing the center of mass of the major body (Venus) that is normal to the inbound asymptotic approach velocity $\vec{v_\infty}\equiv v_\infty\vec{\hat{v}_\infty}$ of the minor body (C/2021 A1), and serves as a useful tool for describing and analyzing such close encounters \citep{kizner1961}. We adopted the commonly used formalism of \citet{valsecchi2003}, which labels the plane by two coordinates $\xi$ and $\zeta$ associated with the orthonormal unit vectors $\vec{\hat{\xi}}$ and $\vec{\hat{\zeta}}$, with Venus at the origin, and with $\vec{\hat{\zeta}}\vec{\times}\vec{\hat{\xi}}=\vec{\hat{v}_\infty}$ rotated such that the projected heliocentric velocity of Venus falls along $-\vec{\hat{\zeta}}$.

\begin{figure}
\centering
\includegraphics[width=\linewidth]{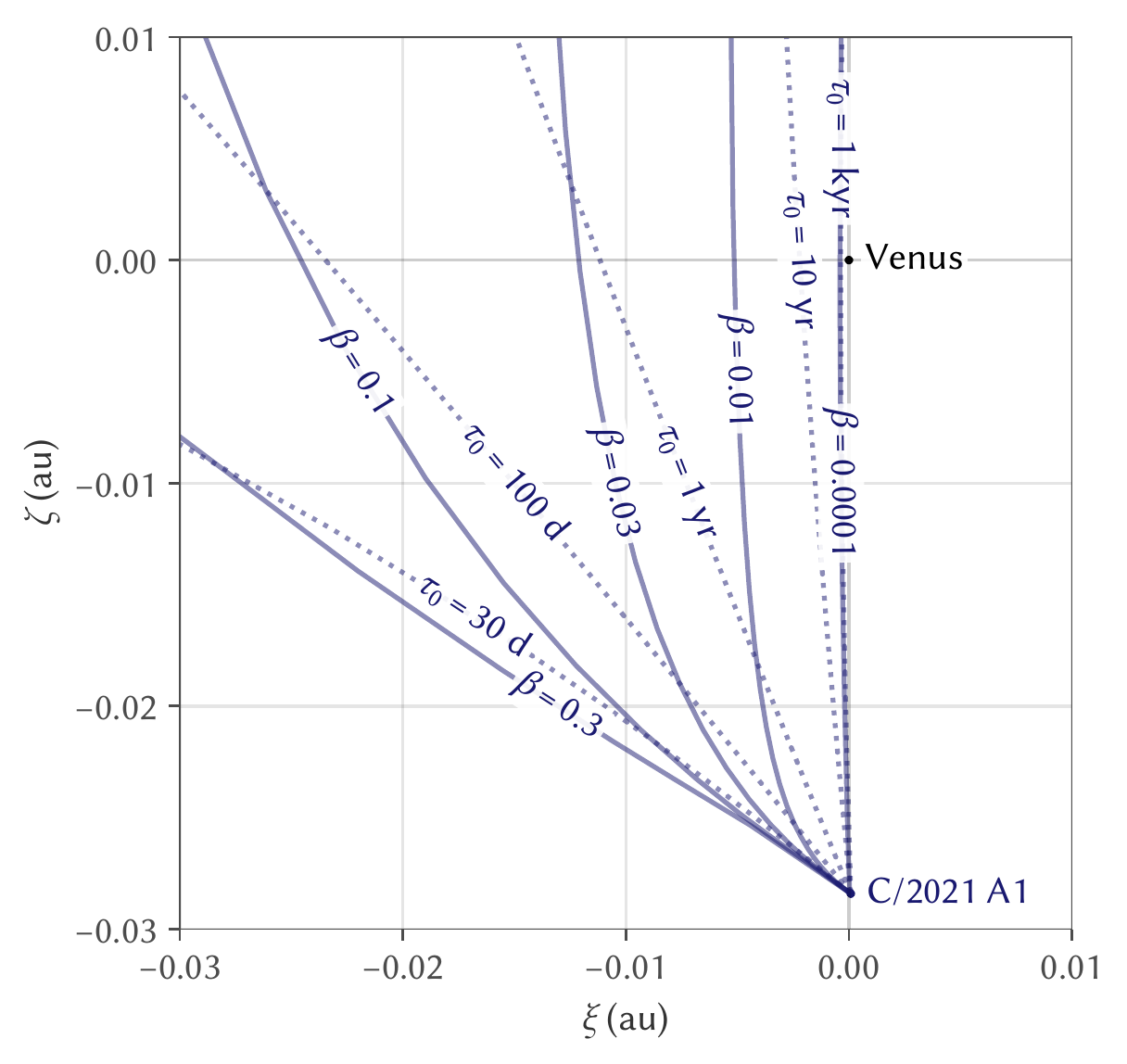}
\caption{Syndynes (solid curves) and synchrones (dotted curves) of C/2021~A1 projected onto the $b$-plane of its encounter with Venus, illustrating the closest approach points of $v_d=0$ dust. Venus falls $\sim$50,000~km from the edge of this dust fan, so a small but nonzero $v_d$ is required for any dust to reach Venus. Note that the labeled dots indicating the positions of Venus and the nucleus of C/2021~A1 are both much larger than the respective objects, for visibility.}
\label{fig:xz}
\end{figure}

Figure~\ref{fig:xz} shows the $b$-plane crossing points for dust of several representative $\beta$ and ejection ages $\tau=\tau_0$ defined from the \edit1{$b$-plane} crossing time of the nucleus $T_0$\edit1{---effectively the time of its closest approach to Venus} (2021~December~18 02:08~UT $\pm~3$~minutes; JPL orbit solution 9). While each dust grain will technically have a different $b$-plane, their relative velocity with respect to each other and to the nucleus is much smaller than their speed with respect to Venus, which ensures that all encounter trajectories are nearly parallel and thus enables the $\xi$ and $\zeta$ of their crossing points to be usefully compared together in this manner.

Additionally, the asymptotic approach speed of $v_\infty=78$~km~s$^{-1}$ of C/2021~A1 to Venus is much faster than the escape speed of 10.4~km~s$^{-1}$ from the surface of Venus, so gravitational deflection of both the comet nucleus and of all dust grains moving with a similar velocity by Venus will be negligible. In this case, the $\xi$ and $\zeta$ where the nucleus and all associated dust grains cross the $b$-plane represents the spatial position of each respective object's closest approach to the planet. All crossing points $\xi^2+\zeta^2\lesssim\tilde{R}_\mathrm{ven}^2$ therefore represent impact on the planet, where we have adopted $\tilde{R}_\mathrm{ven}=6200$~km as the solid radius of Venus, plus a $\sim$100~km buffer as a rough estimate for the portion of atmosphere within which grazing meteors may fully ablate\edit1{, plus an additional $\sim$50~km from gravitational focusing of $v_\infty=78$~km~s$^{-1}$ particles by Venus}.

As Venus falls just outside the $v_d=0$ dust fan by the MOID of $\sim$50,000~km, \edit1{the planet will not encounter an abundance of low $v_d$ dust grains when it crosses the comet's orbital plane and will, instead, only intercept the higher $v_d$ grains that drift to it away from the main fan. To evaluate the plausibility of dust grains reaching Venus, we define a transformation that projects the initial ejection velocity $\vec{v_d}$ for each dust grain of $\beta$ and $\tau_0$ onto its corresponding $b$-plane crossing point $(\xi,\zeta)(\beta,\tau_0,\vec{v_d})$.} This transformation is nearly linear under the linear encounter approximation described above when considering only $(\xi^2+\zeta^2)^{1/2}\ll 0.7$~au (i.e., Venus' orbit radius), and can therefore be described by (\ref{eq:transform}):

\begin{equation}
\label{eq:transform}
\begin{split}
(\xi,\zeta)(\beta,\tau_0,\vec{v_d})&=(\xi,\zeta)(\beta,\tau_0,\vec{0})+\mat{X}_{\beta,\tau_0}\vec{v_d},\\
\mat{X}_{\beta,\tau_0}&\equiv\frac{\partial(\xi,\zeta)}{\partial\vec{v_d}}(\beta,\tau_0,\vec{0})
\end{split}
\end{equation}

Here, $\mat{X}_{\beta,\tau_0}$ is a $2\times3$ Jacobian matrix relating displacements in $\vec{v_d}$ to displacements in $\xi,\zeta$. An isotropic shell of dust ejected at equal speed $v_d$ with a given $\beta$ and $\tau_0$ therefore projects onto a footprint ellipse in the $b$-plane that is centered on the associated $v_d=0$ crossing point, with the semiaxes of the ellipse given by $v_d$ multiplied by the singular values of $\mat{X}_{\beta,\tau_0}$.

\begin{figure}
\centering
\includegraphics[width=\linewidth]{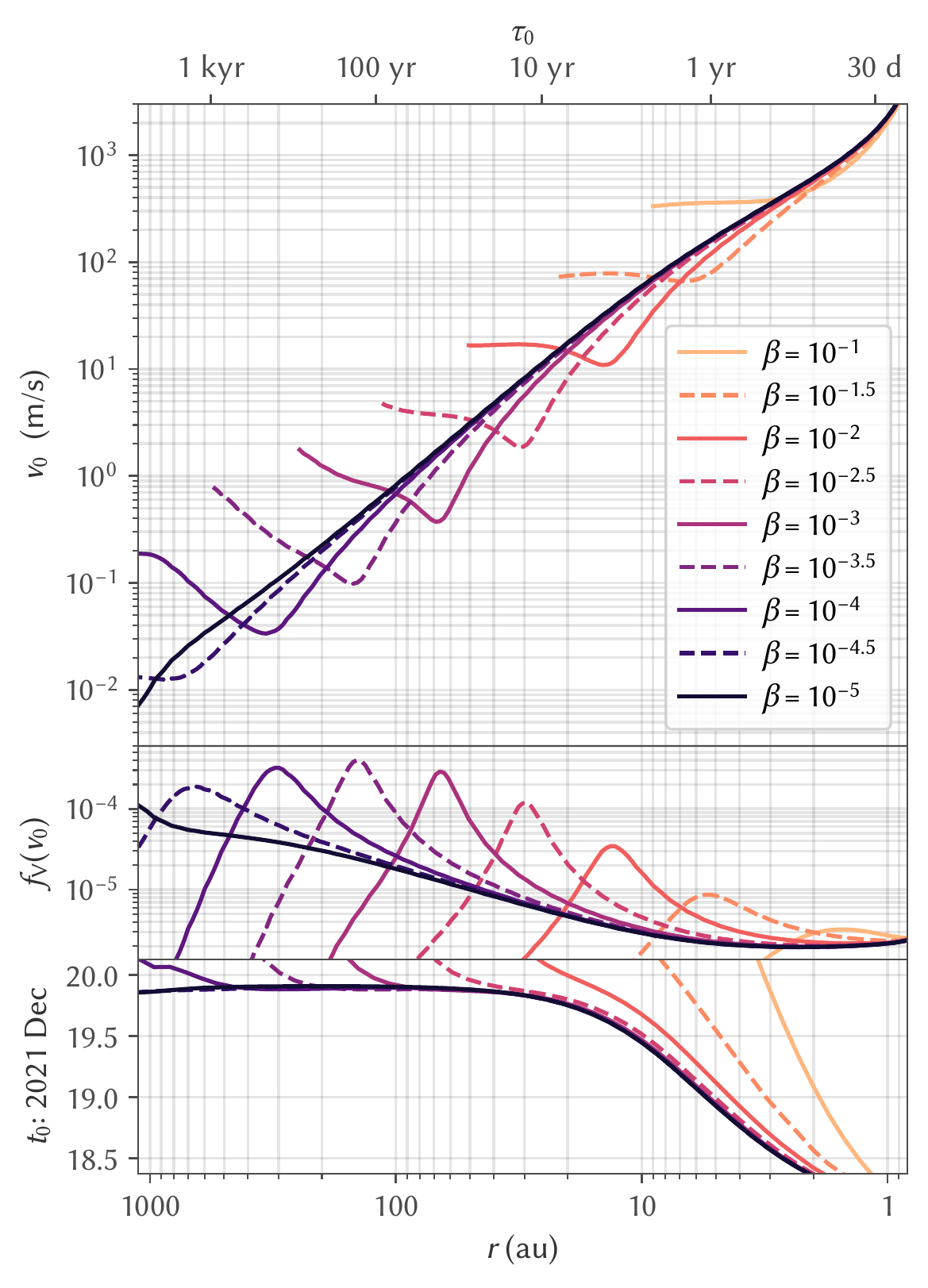}
\caption{Minimum $v_d$ for dust of various $\beta$ ejected at various $r$ to reach Venus ($v_0$), the approximate fraction of dust isotropically ejected with $v_d\sim v_0$ reaching Venus ($f_\mathrm{V}(v_0)$), and the time $t_0$ at which they do so.}
\label{fig:dv}
\end{figure}

For each $\beta$ and $\tau_0$, we compute $(\xi,\zeta)(\beta,\tau_0,\vec{0})$ by two body Keplerian trajectories from the comet's heliocentric osculating orbit at $T_0$ to optimize computing resource usage, and subsequently calculate $\mat{X}_{\beta,\tau_0}$ by finite differences. As $a_\mathrm{in}\approx1900$~au, this approximation is only valid out to $\sim$1000~au before the osculating orbit and the true trajectory substantially diverge. Planetary perturbations will minimally alter the relative trajectory of the nucleus and dust grains, since all relevant trajectories are tightly clustered together compared to interplanetary distances and thus experience nearly the same perturbations. We then follow \citet{farnocchia2014} and use Lagrange multipliers to determine the minimum speed $v_d=v_0$ a dust grain with a given $\beta$ and $\tau_0$ must be ejected with in order to reach Venus, representing a first criterion for any dust grains with a particular set of parameters to be become meteors. The first plot of Figure~\ref{fig:dv} shows that the required $v_0$ generally decreases with increasing $\tau_0$, but with each $\beta$ having an optimum $\tau_0$ where $v_0$ is minimized, corresponding to the $\beta$ and $\tau_0$ combinations with $(\xi,\zeta)(\beta,\tau_0,\vec{0})$ near the origin. These results and the associated approximations were validated by spot checking with full $n$-body simulations that include the gravity of the Sun and eight major planets, which indicate the computed $\vec{v_0}$ are generally accurate to within $\sim$10\% for $r(\tau_0)\lesssim30$~au, and within a factor of about two for $r(\tau_0)\lesssim1000$~au.

While grains ejected at $v_d>v_0$ in certain directions can reach Venus, only a small fraction of ejected grains will actually do so. This fraction depends on the size of the footprint ellipse, with a larger footprint generally translating to a smaller fraction of dust reaching Venus, once $v_d>v_0$ is met. We consider a meteor production efficiency $f_\mathrm{V}(v_0)$ defined as the ratio of $\pi\tilde{R}_\mathrm{ven}^2$ to the area of the footprint ellipse for $v_d=v_0$, which very roughly corresponds to the fraction of dust ejected isotropically at $v_d\sim v_0$ that will impact Venus. The second plot of Figure~\ref{fig:dv} shows that $f_\mathrm{V}(v_0)$ is a maximum of $\sim$0.03\% for this encounter for dust grains of $\beta\sim10^{-3}$--$10^{-4}$ ejected at $r(\tau_0)\sim70$--300~au with $v_d\sim v_0\sim10^{-1}$~m~s$^{-1}$---comparable to the escape speed from the surface of a kilometer-scale nucleus. Larger and older grains will generally have even lower $v_0$, but which correspond to larger $b$-plane footprints and thus lower $f_\mathrm{V}(v_0)$.

The last plot of Figure~\ref{fig:dv} shows the impact time $t_0$ for $v_d=v_0$ dust grains. Meteors associated with distant activity ($\beta\lesssim10^{-2.5}$, $r(\tau_0)\gtrsim30$~au)---where efficiency $f_\mathrm{V}(v_0)>0.01\%$---will largely be concentrated to within $\sim$1~h of 2021~December~19 21:00 UT, when Venus passes through its MOID point with the comet's heliocentric orbit. \edit1{Any meteors resulting from more recent dust production may begin as soon as $\sim$1.5~d earlier near $T_0$.}

\edit1{However, no meteors will occur at any time if the comet does not actually produce any dust of $v_d\gtrsim v_0$ on its approach.} Dust produced since the comet's earliest observations at $r=7.5$~au require a very high $v_d>v_0>100$~m~s$^{-1}$ to reach Venus. A $\beta=10^{-2}$ dust grain---near the average observed from C/2021~A1---ejected at $v_d\sim100$~m~s$^{-1}$ in the sunward direction while at $r\sim5$--10~au would experience radiation pressure acceleration relative to the nucleus of ${\sim}10^{-6}$~m~s$^{-2}$ in the anti-sunward direction. The grain would not be turned around and pushed into a tail for $>$1~yr while ${\gtrsim}10^6$~km from the nucleus. The observed dust coma, however, is two orders of magnitude smaller, yet is already folded into a tail, so we do not observe dust grains ejected near this speed.

Aside from the cases of an extremely large nucleus \citep{jewitt1999} and near-Sun comets well within the orbit of Venus \citep{jones2018}---neither of which apply to C/2021~A1---dust of the observed $\beta\lesssim10^{-2}$ and needed $v_d>100$~m~s$^{-1}$ are only produced by explosive outbursts where the gas is of sufficient density to accelerate dust to speeds comparable to the outflow speed of up to $\sim$1~km~s$^{-1}$. Such high speed outbursts and the distinctive dust shells morphology they produce are occasionally observed from from Jupiter family comets like 15P/Finlay at $r\sim1$~au \citep{ye2015} and 17P/Holmes at 2.4~au \citep{russo2008}, as well as from active centaurs like 29P/Schwassmann--Wachmann \citep{trigo2008} and 174P/Echeclus \citep{rousselot2008} at $r>5$~au. One such outburst was also observed from the long period comet C/2012~X1 (LINEAR) at 2.7~au \citep{miles2013}. A dust shell produced by such an outburst from C/2021~A1 expanding at $v_d\gtrsim0.5$~km~s$^{-1}$ can reach Venus if the outburst occurs before the comet reaches $r\approx2$~au in 2021~September. However, given the exceptional rarity of high speed dust outbursts from long period comets, such an event appears highly unlikely for C/2021~A1. Therefore, neither the comet's observed past activity nor its future activity are likely to yield any meteors on Venus, and the most likely source for meteors on Venus becomes dust producing activity on the comet from before it was first observed.

Additionally, the comet's large $a_\mathrm{in}\approx1900$~au implies that perihelion activity from previous apparitions will not substantially affect the meteoroid flux at Venus: Even without non-gravitational forces (i.e., \edit1{$\beta\to0$}), particles separated from the nucleus with a modest $v_d=1$~m~s$^{-1}$ at the previous perihelion will be widely spaced along the orbit, preceding or trailing the nucleus by up to $\sim$30~kyr in the current apparition. We therefore expect no substantial enhancement in the meteoroid flux near the nucleus from activity on previous apparitions. Any meteor activity observed on Venus from C/2021~A1 would thus most likely originate from distant activity by the comet from earlier in the present apparition.

\subsection{Distant Dust Production}

Pre-observational distant activity on C/2013~A1 at $r>10$~au was inferred from the meteor shower it produced on Mars. In that encounter, Mars lay on the $b$-plane at the intersection of the $\beta=1.43\times10^{-4}$ syndyne and the $r(\tau_0)=22.5$~au synchrone, and thus efficiently intercepted dust grains produced with these and similar properties \citep{farnocchia2014}. While the earliest analyses by \citet{vaubaillon2014} and \citet{moorhead2014} had suggested that an abundance of younger, high velocity grains could potentially dominate the meteoroid flux at Mars, they required comet behavior ultimately inconsistent with the observed dust production of the comet. \citet{ye2014}, \citet{tricarico2014}, and \citet{li2014} directly detected and thus constrained the quantity and velocity of $\beta\lesssim10^{-2}$ grains produced by the comet, showing that the small and recently produced grains were far too slow to reach Mars. \citet{kelley2014} likewise simulated dust production starting from $r=13$~au, showing that there would be negligible meteor activity from dust produced within this distance.

The remaining possibility of a substantial level of large grains of $\beta\sim10^{-3}$--$10^{-4}$, or $a_d\sim1$--10~mm, being produced at $r>10$~au had not been seriously considered prior to the encounter, so the MAVEN IUVS detection of a temporary metallic vapor layer lasting $\sim$2~d with a strength consistent with ${\sim}10^4$~kg of ablated meteoroids from the comet \citep{schneider2015}---far in excess of the layer's quiescent level \citep{crismani2017}---was largely unexpected. Those observations were supplemented by detections of an apparent influx of ions---particularly metallic ions likely of meteoritic origin---into the planet's ionosphere by instruments aboard MAVEN \citep{benna2015}, Mars Reconnaissance Orbiter \citep{restano2015}, and Mars Express \citep{sanchezcano2020}.

We note that \citet{schneider2015} finds that the high $\sim$120~km altitude of the metallic vapor layer---and thus of meteor ablation---to be consistent with predominantly $a_d\sim1$--100~$\mu$m meteoroids of $\rho_d<1$~g~cm$^{-3}$, which they suggest implies intercepted grains were actually recently ejected from the nucleus at very high $v_d$, rather than the expected 1--10~mm grains from $r>10$~au. However, ${\sim}10^4$~kg of grains of $a_d\sim1$--100~$\mu$m with typical $\rho_d\sim0.3$--1~g~cm$^{-3}$ and $\sim$4\% albedo spread over the cross sectional area of Mars would have had a readily observable $V$-band surface brightness of $\sim$21--27~mag~arcsec$^{-2}$, while telescopic observations of the comet during the encounter were consistent with pre-encounter observations, and revealed no such high $v_d$ dust \citep{li2016}. Therefore, the intercepted grains were most likely the expected 1--10~mm grains from $r>10$~au, and we speculate that the high ablation altitude may instead reflect a weaker grain structure that is more prone to fragmentation than exhibited by the much older meteoroids in Earth-crossing streams---which traditional meteor models are largely based on---as such fragile grains in these multi-apparition streams may have already broken down prior to reaching Earth.

Such distant dust activity may not be unusual. More recently, the large Oort cloud comet C/2017~K2 (PANSTARRS) was directly observed to be producing millimeter-sized grains at $r>20$~au \citep{hui2017}. This activity appears consistent with surface sublimation of CO \citep{meech2017,jewitt2019,yang2021}, a process that can be thermally efficient out to $r\sim120$~au \citep{meech2004}. \citet{jewitt2021} concluded from monitoring of the comet's dust morphology that the observed dust activity was underway out at least to $r\approx35$~au, finding a dust production rate following

\begin{equation}
\label{eq:m}
\dot{m}\approx\dot{m}_{10}\times(10~\mathrm{au}/r)^2
\end{equation}

with $\dot{m}_{10}\sim1000$~kg~s$^{-1}$. They also derive a \edit1{model for the mean dust ejection speed}

\begin{equation}
\label{eq:vd}
\langle v_d\rangle\approx v_{10}\times\beta^{0.5}\times(10~\mathrm{au}/r)
\end{equation}

and measured $v_{10}\approx170$~m~s$^{-1}$ for C/2017~K2, a value consistent with a $R_\mathrm{nuc}\approx6$~km radius nucleus covered in CO ice, following a theoretical $v_{10}\propto R_\mathrm{nuc}^{1/2}$ relation.

C/2021~A1 is intrinsically much fainter than C/2017~K2, and is thus likely to be much smaller, have a much less volatile surface, or both. The dust-contaminated magnitudes reported in the earliest pre-discovery observations \citep{leonard2021a} set only a very weak upper bound of $R_\mathrm{nuc}\lesssim7$~km, assuming a typical $\sim$4\% albedo. Additionally, the barycentric $a_\mathrm{in}\approx1900$~au suggests the comet has previously survived a perihelion passage of a similar $q\approx0.6$~au to its present apparition at least once before \citep{krolikowska2017}, while dynamically new Oort cloud comets of $R_\mathrm{nuc}\lesssim0.5$~km frequently disintegrate at comparable or larger $r$ \citep[e.g.,][]{farnham2001,li2015,combi2019}. Although we lack a rigorous statistical analysis of survival rate by nucleus size, the survival of C/2021~A1 crudely suggests $R_\mathrm{nuc}\gtrsim0.5$~km, and we adopt $R_\mathrm{nuc}\approx1$~km as the expected value, reflecting both the lower limit and the bottom heavy size distribution of comet nuclei \citep{boe2019}. This assumed size is additionally comparable to the lower bound of $R_\mathrm{nuc}>0.9$~km obtained by \citet{schleicher2002} for C/2000~WM$_1$ (LINEAR), a comet with very similar $a_\mathrm{in}\approx1900$~au and $q\approx0.6$~au to C/2021~A1, as well as a similar brightness at $r\sim5$~au pre-perihelion.

A nucleus of $R_\mathrm{nuc}\approx1$~km gives $v_{10}\approx70$~m~s$^{-1}$ under \citet{jewitt2021}'s CO surface sublimation model. This $v_{10}$ implies that a grain of any $\beta$ ejected at any $r$ in the sunward direction will be turned around by radiation pressure at a distance of $d_\mathrm{ta}\sim80{,}000$~km away from the nucleus, setting a minimum size for the coma. Figure~\ref{fig:tail}, however, shows a much smaller coma with a sunward edge only ${\sim}3''$ from the nucleus in the first epoch, which translates to $d_\mathrm{ta}\sim2000$~km after correcting for the $\alpha\approx11^\circ$ by approximating the coma as a paraboloid. The small $d_\mathrm{ta}\sim2000$~km corresponds to a much lower $v_{10}\sim10$~m~s$^{-1}$, so the nucleus evidently does not contain large quantities of actively sublimating CO as that of C/2017~K2 appears to have. For comparison, C/2013~A1 was observed to have $v_{10}\sim20$~m~s$^{-1}$ at similar $r$ \citep{ye2014}. We note, however, that $v_{10}$ may have been higher in the past if the $r\sim5$~au activity is being driven by a less volatile substance like CO$_2$ or H$_2$O, or if the surface only became depleted of CO earlier during its present approach to the Sun. The latter could result from seasonal effects if a portion of the nucleus remained shaded through the outbound leg of its previous apparition, and was thus still freshly resurfaced from perihelion until efficient CO sublimation began at $r\sim120$~au. We therefore regard $v_{10}\approx70$~m~s$^{-1}$ as a plausible albeit unlikely upper limit for dust produced by quiescent surface sublimation.

\begin{figure}
\centering
\includegraphics[width=\linewidth]{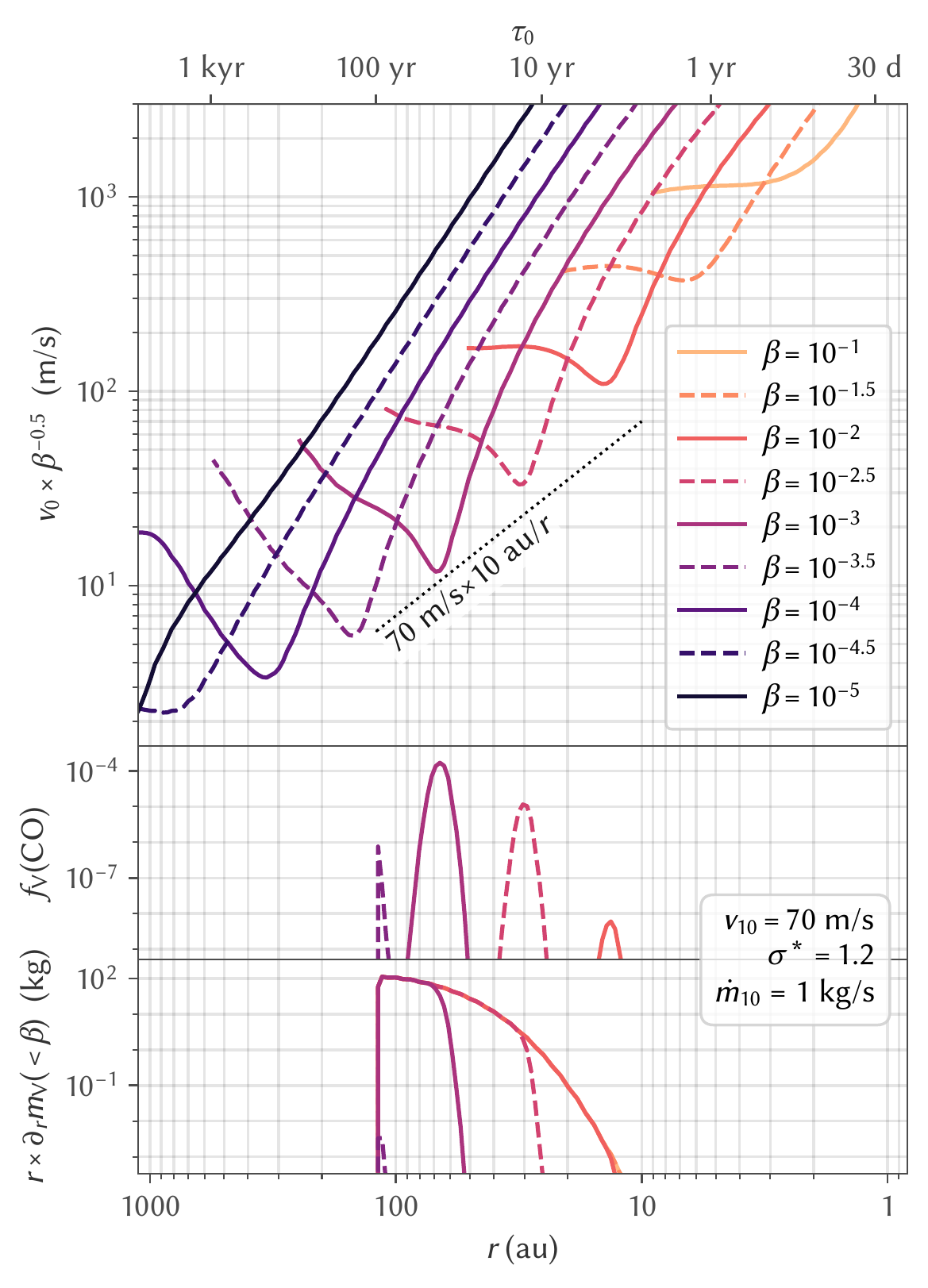}
\caption{The $v_0$ from Figure~\ref{fig:dv} scaled by $\beta^{-0.5}$ relative to the $v_{10}=70$~m~s$^{-1}$ for a 2~km diameter nucleus with CO surface sublimation, the fraction $f_{\mathrm{V}}(\mathrm{CO})$ of such dust intercepted by Venus, and the intercepted mass $m_\mathrm{V}$ below the indicated $\beta$ produced per unit interval of $\ln r$---denoted by $r\times\partial_r m_\mathrm{V}({<}\beta)$---serving as a rough approximation for the intercepted mass produced near $r$.}
\label{fig:dvco}
\end{figure}

The first plot of Figure~\ref{fig:dvco} modifies Figure~\ref{fig:dv} by scaling the calculated $v_0$ by a factor $\beta^{-0.5}$ to simultaneously compare the necessary $v_0$ for dust grains of different sizes, since the quantity $v_d\times\beta^{-0.5}$ is approximately independent of $\beta$ under the model in (\ref{eq:vd}). The plot shows that the $v_{10}\approx70$~m~s$^{-1}$ of a CO-covered surface at $r<120$~au is nearly the minimum required for dust of any size released by the comet at any point on its current apparition to have the needed $v_d\geq v_0$ to reach Venus. The second plot shows the corresponding $f_\mathrm{V}$ for the footprint ellipse set by $v_{10}=70$~m~s$^{-1}$, which is a rough approximation of the dust ejected under this model of a given $\tau_0$ and $\beta$ that impact Venus. We also consider $v_d\times\beta^{-0.5}$ to be lognormally distributed about the nominal value, with a geometric standard deviation $\sigma^*$, and select a plausible $\sigma^*=1.2$ for this figure. The true $\sigma^*$ is affected by many factors, including the distribution of particle drag coefficients and nucleus non-uniformity, and a precise computation is beyond the scope of this analysis. We also consider a dust mass production rate of $\dot{m}_{10}=1$~kg~s$^{-1}$ with (\ref{eq:m}) for the third plot, which shows the mass $m_\mathrm{V}({<}\beta)$ of all dust grains larger than the indicated $\beta$ reaching Venus under this model produced per unit interval of $\ln r$, denoted by $r\times\partial_r m_\mathrm{V}({<}\beta)$. The plot shows that these meteoroids intercepted by Venus will be principally by $\beta\sim10^{-3}$ grains produced near $r\sim100$~au around the maximum $r$ where efficient CO sublimation is possible.

\begin{figure}
\centering
\includegraphics[width=\linewidth]{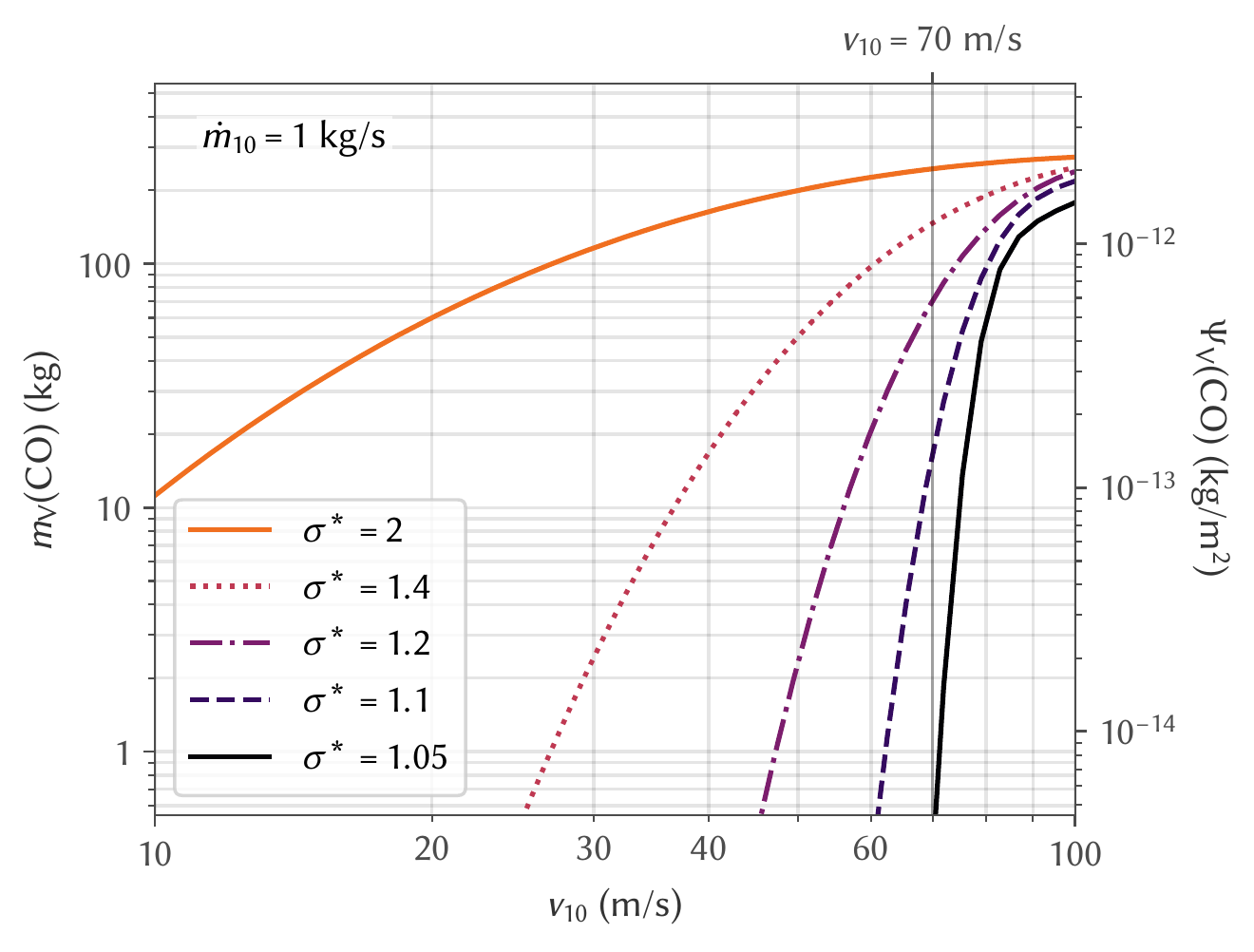}
\caption{Total mass $m_\mathrm{V}(\mathrm{CO})$ and mass flux density $\Psi_\mathrm{V}(\mathrm{CO})$ intercepted by Venus as as functions of typical normalized dust ejection speed $v_{10}$ and geometric dust speed dispersion $\sigma^*$ for isotropic dust production $\dot{m}\propto r^{-2}$ and $\dot{m}_{10}\equiv\dot{m}(10~\mathrm{au})=1$~kg~s$^{-1}$ from the CO sublimation limit $r=120$~au to 10~au.}
\label{fig:mco}
\end{figure}

As $v_{10}=70$~m~s$^{-1}$ is very near the threshold speed for any dust to reach Venus, the cumulative mass of meteoroids intercepted by Venus $m_\mathrm{V}(\mathrm{CO})$ under this CO sublimation model is extremely sensitive to the precise values of $v_{10}$ and $\sigma^*$. Figure~\ref{fig:mco} plots the intercepted mass $m_\mathrm{V}(\mathrm{CO})$ and corresponding mass flux $m_\mathrm{V}(\mathrm{CO})$ for a variety of $v_{10}$ and $\sigma^*$ for the adopted $\dot{m}_{10}=1$~kg~s$^{-1}$. A substantial meteoroid flux at Venus occurs only for high $v_{10}\gtrsim70$~m~s$^{-1}$ or high $\sigma^*\gtrsim1.2$, which will produce $m_\mathrm{V}(\mathrm{CO})\sim10^2$~kg of meteors on Venus. We note that the model treats dust as non-interacting particles that can be linearly superposed, so the results may be scaled to any $\dot{m}_{10}$ by $m_\mathrm{V}(\mathrm{CO})\propto\dot{m}_{10}$. However, if the $v_{10}\sim10$~m~s$^{-1}$ observed near $r\sim5$~au held throughout its approach, the comet will produce negligible meteor activity on Venus without an exceptionally high $\sigma^*\gtrsim2$ where a substantial fraction of dust is ejected at several times the nominal speed. In contrast, an equivalent model for C/2013~A1's encounter with Mars---where Mars actually passed within the $v_d=0$ dust fan---gives the observed ${\sim}10^4$~kg of meteors for the adopted $\dot{m}_{10}\sim1$~kg~s$^{-1}$ at all reasonable $v_{10}$ and $\sigma^*$, indicating that CO surface sublimation adequately explains the meteor activity on Mars from C/2013~A1.

Distant activity may also be driven by mechanisms other than the quiescent surface sublimation we have modeled. For example, \citet{sekanina2014} \edit1{note} that C/2012~S1 was trailed by a stream of dust extending several degrees behind the nucleus, which syndyne--synchrone analysis indicates were comprised of $\beta\lesssim10^{-3}$ grains ejected at $r\gtrsim10$~au, and suggests based on the sublimation behavior that it may have contained pebbles of $\beta\sim10^{-4}$ from as far as $r\sim100$~au produced by annealing of amorphous ice. \citet{disanti2016} and \citet{feldman2018} found the comet had a low $Q(\mathrm{CO})/Q(\mathrm{H}_2\mathrm{O})\sim1.5\%$ near $r\sim1$~au, which suggests that CO surface sublimation was unlikely to have been responsible for the distant activity, unless the surface abundance of CO was considerably higher earlier.

Extremely distant activity has also been inferred from the trajectories of fragmented comets. The comet pair C/1988~F1 (Levy) and C/1988~J1 (Holt) were discovered following nearly the same orbit spaced only 76~d apart---too close for the progenitor to have split on its previous apparition, yet too far to have split after entering the planetary region on its current apparition \citep{marsden1988}. Simulations by \citet{sekanina2016} suggest the pair likely separated at $\sim$1~m~s$^{-1}$ centuries earlier at $r\sim300$--900~au, already on its approach to perihelion. Likewise, the arrival clustering of the Kreutz sungrazing comet family appears to require its members to split decades before perihelion at $r\gtrsim50$~au and $\sim$10~m~s$^{-1}$ \citep{sekanina2004}.

The physical mechanisms behind the distant fragmentation events are unclear, and no obvious large fragments or companions to C/2021~A1 have been discovered as of 2021~May. However, if such mechanisms are present that can split comets several tens or even hundreds of astronomical units before perihelion at $\sim$1--10~m~s$^{-1}$, it seems plausible that they may, under other circumstances that may or may not apply to C/2021~A1, propel smaller, unseen fragments or large dust grains from the comet at similar $r$ with the $v_d\geq v_0\sim0.1$--1~m~s$^{-1}$ necessary to reach Venus. Without a clear physical explanation for this behavior, however, we cannot meaningfully speculate on the likelihood of such activity contributing substantially to the meteoroid flux at Venus.

\subsection{Meteor Observability}

\begin{figure}
\centering
\includegraphics[width=\linewidth]{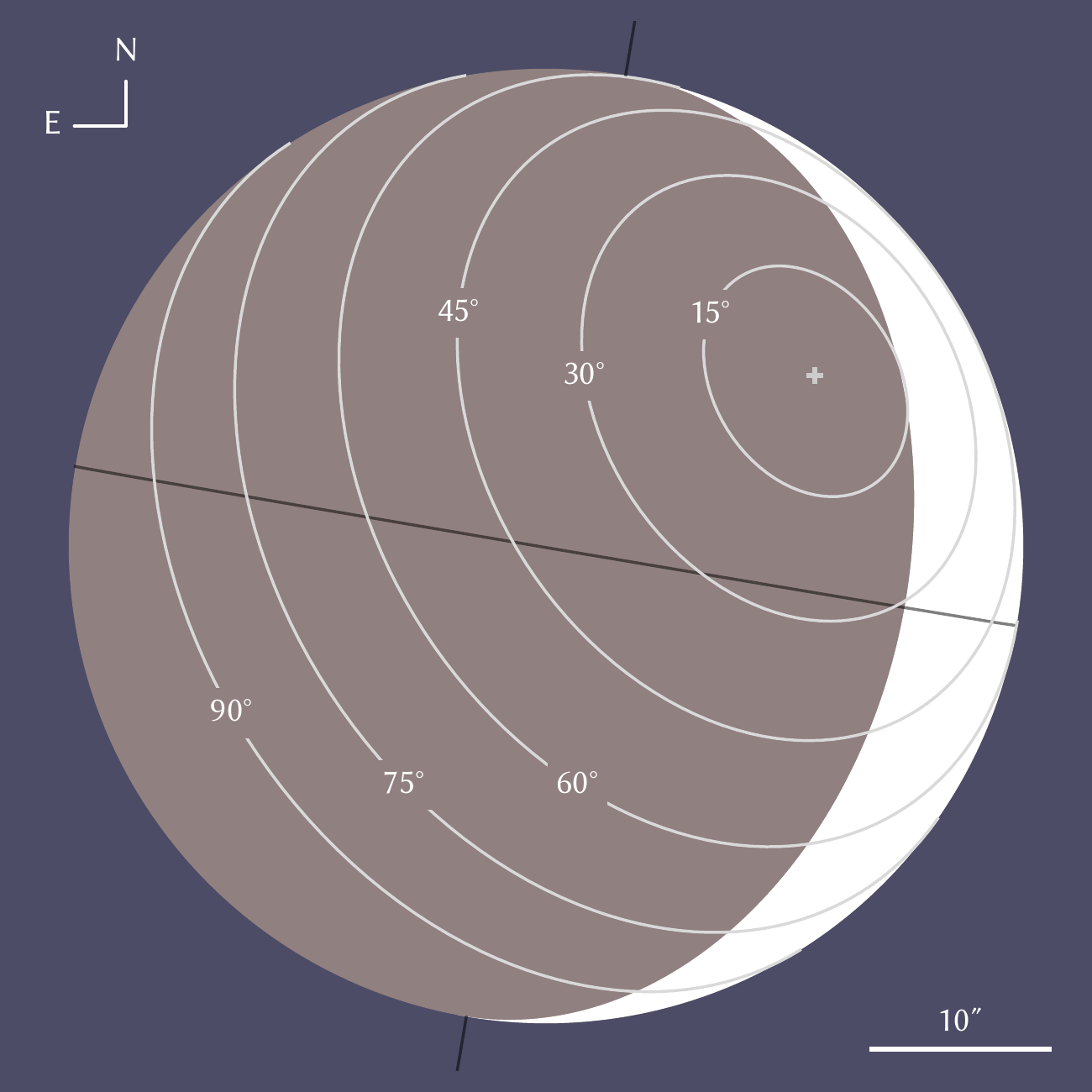}
\caption{Spatial map for potential meteors on Venus viewed from Earth at 2021~December~19 21:00 UT: Contours indicate curves of constant radiant zenith angle, and encircle the subradiant point (marked by ``+''). Dark lines indicate Venus' equator and axis of rotation, while the white crescent represents the solar illuminated portion of the planet. Meteors are possible anywhere on the disk the zenith angle is ${<}90^\circ$, but are most probable near the subradiant point.}
\label{fig:venus}
\end{figure}

Venus' MOID \edit1{point passage} near 2021~December~19 21:00~UT, when \edit1{large,} distantly released meteoroids from C/2021~A1 are most favored to reach the planet, will occur while Earth is only 0.32~au away. From Earth, Venus will have an apparent diameter of $53''$ at $28^\circ$ solar elongation with a thin crescent illuminated at $140^\circ$ phase angle. Meteors from the comet will \edit1{approach the planet} at $v_\infty=78$~km~s$^{-1}$ from \edit1{directions tightly clustered within a few arcminutes of the radiant at} R.A. $10^\mathrm{h}52^\mathrm{m}$ and decl. $+36^\circ55'$ \edit1{(J2000), prior to gravitational focusing}. This \edit1{radiant} is overhead from a subradiant point at longitude $353^\circ$ and latitude $+27^\circ$, which is on both the Earth-facing and nightside hemispheres near the evening terminator. Figure~\ref{fig:venus} plots contours of radiant zenith angle over the apparent disk and illuminated crescent of Venus viewed from Earth, indicating the portion of the disk where meteors from the comet, if any, are likely to appear.

At a distance of 0.32~au, meteors on Venus will be $\sim$28~mag fainter than a terrestrial meteor of the same intrinsic brightness observed from a typical $\sim$100~km away. A 0.5~m aperture telescope and detector with a system throughput of 30\% under ideal sky conditions is sensitive to point sources out to $V\sim16$ in exposures as short as $\sim$0.1~s, corresponding to \edit1{Venus} meteors of $V\sim-12$ \edit1{from $\sim$100~km away}. While terrestrial meteors from solar system comets are never as fast as the $v_\infty=78$~km~s$^{-1}$ expected for C/2021~A1 meteors on Venus, Leonid meteors are nearly so, at $v_\infty=71$~km~s$^{-1}$, so serve as reasonably close analogs, although C/2021~A1 meteors of the same \edit1{mass} will be $\sim$20\% more energetic and additionally brighter from the shorter scale height in Venus' upper atmosphere \citep{christou2004,mcauliffe2006}. Leonids of $V\sim-12$ are not uncommon and may be produced by meteoroids of $\sim$0.2~kg in mass, which have $\beta\sim10^{-5}$ \citep{shrbeny2009}. No dust with $\beta\lesssim10^{-3.5}$ produced by surface sublimation of CO can reach Venus, so other mechanisms---such as those responsible for the distant pre-perihelion splitting of C/1988~F1 and C/1988~J1, and of the Kreutz sungrazers, as discussed in the previous section---are required to produce any meteors bright enough to be observed on Venus from Earth.

In practice, $V\sim16$ serves as an optimistic limit given that any meteors must be observed in close proximity to Venus' brilliantly illuminated dayside, and coronagraphic optics like those developed to search for lightning on the nightside may be required to approach this limit \citep[e.g.,][]{hansell1995}. Bandpass filters encompassing select bright meteor emission features, such as the Na~I D lines at 589.0/589.6~nm or the O~I line at 777~nm \citep{carbary2004}, may further aid in stray light and background rejection. Additionally, the low elongation of Venus means any observations will be limited to high airmass in twilight, while the short window of no more than $\sim$2~h for bright meteors associated with distant activity geographically restricts any such observations to a narrow corridor through the Atlantic Ocean and the easternmost portions of Canada and Brazil.

\edit1{Fluorescence} from a meteoritic metallic vapor layer, like that detected by \citet{schneider2015} on Mars following the C/2013~A1 encounter, may also be observable near the terminator---particularly near the northern limb, where apparent column density is elevated by foreshortening. One appealing candidate is Na~I, whose D lines are readily observable from the ground, and for which a persistent meteoritic layer of $10^{13}$--$10^{14}$~atoms~m$^{-2}$ is well-observed over Earth \citep[e.g,][]{cabannes1938,chapman1939,chamberlain1956}. As on Earth, Venus' low heliocentric radial velocity of $-0.2$~km~s$^{-1}$ during the encounter combined with the deep Na~I D Fraunhofer absorption lines in the solar spectrum strongly suppresses the intensity of the resonance emission, with a fluorescence efficiency of $\sim$1~photon~s$^{-1}$~atom$^{-1}$ for the D2 (589.0~nm) line. Keck HIRES spectroscopy has been demonstrated to be sensitive to emission lines on Venus down to a vertical equivalent $\sim$20~R \citep{slanger2001}, and other large telescopes equipped with echelle or Fabry--P\'{e}rot etalon spectrographs with spectral resolving power $\gtrsim$50,000 that can cleanly distinguish Venusian emission lines from telluric lines---separated by 8.3~km~s$^{-1}$ in radial velocity---may be similarly sensitive. This limit translates to ${\sim}2\times10^{11}$~atoms~m$^{-2}$ of Na~I, or the typical Na content of $\sim$200~kg of ablated meteoritic material distributed over one hemisphere of Venus in the absence of ionization or other upper atmospheric chemical reactions that may remove Na~I. We note, however, that Venus lacks a persistent Na~I layer like Earth's, suggesting that Na~I is indeed rapidly consumed by atmospheric chemistry \citep{krasnopolsky1983}. A more careful analysis is required to evaluate if meteoritic Na~I can sufficiently persist for $\gtrsim$1~h (i.e., at least \edit1{through the end} of the meteor shower) for any plausible mass of ablated meteoritic material to be detected in this manner.

Additionally, the Akatsuki orbiter remains in operation around Venus \citep{nakamura2011}, and will likely be the only spacecraft in the immediate vicinity of the planet during the C/2021~A1 encounter; no flybys of Venus are planned for Parker Solar Probe\footnote{\url{http://parkersolarprobe.jhuapl.edu/The-Mission/index.php\#Timeline}}, Solar Orbiter\footnote{\url{https://sci.esa.int/web/solar-orbiter/-/44181-mission-operations}}, or BepiColombo\footnote{\url{https://sci.esa.int/web/bepicolombo/-/48871-getting-to-mercury}} within one month of the event. Absent an extreme level of distant activity by unknown mechanisms or severe violations of our defined assumptions, the encounter poses negligible risk to spacecraft health. Even an enormous $\dot{m}_{10}\sim1000$~kg from CO surface sublimation comparable to that of C/2017~K2 and ${\sim}1000\times$ larger than inferred for C/2013~A1 will produce a meteoroid fluence of ${\lesssim}10^{-3}$~particles~m$^{-2}$ at Venus for all plausible values of $v_{10}$ and $\sigma^*$.

Akatsuki, however, may have the capability to observe any meteors and their subsequent effects on Venus. Of note, Akatsuki carries a Lightning and Airglow Camera \citep[LAC;][]{takahashi2018} that monitors for lightning at the 777~nm O~I line that is also a strong meteor emission feature, and thus should be sensitive to bright meteors. However, LAC is only operational for 20--30~minutes per 10~d orbit when the Sun is fully eclipsed by Venus from the spacecraft, so the camera is unlikely to be operable during the brief window of possible meteor activity from C/2021~A1. Akatsuki also carries an Ultraviolet Imager \citep[UVI;][]{yamazaki2018} which includes a 283~nm channel covering the Mg~II resonance lines at 279.6/280.3~nm that are among the brightest emission features of terrestrial meteors \citep{carbary2004} and were also the brightest lines observed in the C/2013~A1 meteoritic layer on Mars \citep{schneider2015}. UVI can integrate for up to 11~s per exposure, and may be useful for directly imaging bright meteors during the encounter as well as to constrain the brightness of any Mg~II layer above the twilit limb afterward, although the wide 14~nm bandpass means such observations will likely be strongly impacted by stray light originating from the illuminated dayside if not performed in eclipse. Finally, features in the ionospheric electron density profile have been characterized as meteoritic ion layers \citep{patzold2009,withers2013}, and Akatsuki's Radio Science \citep[RS;][]{imamura2017} radio occultation experiment may be sensitive to large perturbations to these layers comparable to that observed by similar methods during C/2013~A1's encounter with Mars \citep{restano2015,sanchezcano2020}.

\section{Conclusions}

The upcoming encounter of C/2021~A1 to Venus is among the closest known of a long period comet to any planet, and is surpassed in recent history only by the 2014 encounter of C/2013~A1 to Mars, which produced a meteor shower on the planet that was indirectly observed by multiple spacecraft. To help ascertain the potential for meteors on Venus, we collected imagery and spectroscopy of C/2021~A1 following its discovery, in advance of the encounter. We found the comet's dust color and size to be typical of other dynamically old long period comets at its distance, with the tail optically dominated by $a_d\sim0.1$--1~mm grains produced within the prior year. Our nondetection of CN sets an upper limit of CN/$Af\rho<10^{22.5}$~molec~s$^{-1}$~m$^{-1}$ that is considerably lower than values typically reported; however, few such measurements have been reported at the $r=4$--5~au of our observations, so the low CN/$Af\rho$ may not necessarily be abnormal for comets at this distance.

Trajectory analysis shows that---barring an unlikely $v_d\gtrsim0.5$~km~s$^{-1}$ outburst before the comet reaches $r\approx2$~au in 2021~September---large dust grains of $a_d\gtrsim1$~mm released at $r\gtrsim30$~au are dynamically most strongly favored to reach Venus, but cometary dust production behavior at such large $r$ is poorly constrained. These large grains inefficiently scatter light, and their presence in the dust trail cannot be usefully constrained telescopically. Quiescent surface sublimation of CO beginning as far as $r\sim120$~au can plausibly produce $a_d\sim1$~mm dust grains that become meteors on Venus, but likely requires the nucleus to have had a much greater abundance of near-surface CO than the minute coma we observed suggests is presently accessible. Other distant activity mechanisms, such as those responsible for splitting comet nuclei hundreds of astronomical units away, could potentially produce larger meteors on Venus, the brightest of which may be directly observable from Earth \edit1{around 2021~December~19 20--22:00 UT}. As with C/2013~A1's Mars encounter, the aftermath of a particularly strong meteor shower can also be detected through the formation of a meteoritic metallic vapor layer and through perturbations to the ionosphere, from Earth and/or by the Akatsuki orbiter around Venus. We encourage observations of this event to take advantage of the rare opportunity to probe an otherwise unseen dust trail, and subsequently constrain cometary behavior at distances where no comets have ever been observed.

\bigskip 
We thank Michael S. P. Kelley and Lori M. Feaga for help with collecting observations, Ludmilla Kolokolova for discussions on the properties of cometary dust grains, and an anonymous referee for helpful comments and suggestions in their review of this manuscript. We additionally thank Joel Pearman, Kevin Rykoski, and Carolyn Heffner for observing support with the Palomar Hale Telescope, as well as Sydney Perez and Ana Hayslip for their support with the Lowell Discovery Telescope.

This research makes use of observations from the Hale Telescope at Palomar Observatory, which is owned and operated by Caltech and administered by Caltech Optical Observatories.

These results made use of the Lowell Discovery Telescope (LDT) at Lowell Observatory. Lowell is a private, non-profit institution dedicated to astrophysical research and public appreciation of astronomy and operates the LDT in partnership with Boston University, the University of Maryland, the University of Toledo, Northern Arizona University and Yale University. The University of Maryland observing team consisted of Quanzhi Ye, James Bauer, Michaela Blain, Adeline Gicquel-Brodtke, Tony Farnham, Lori Feaga, Michael Kelley, and Jessica Sunshine. 

This research has made use of data and/or services provided by the International Astronomical Union's Minor Planet Center.

This work was supported by NSF award AST1852589. S.~V. is supported by an NSF Graduate Research Fellowship and the Paul \& Daisy Soros Fellowship for New Americans.

\facilities{DCT (LMI), Hale (DBSP, WIRC)}

\software{Astropy \citep{astropy2013}, emcee \citep{foreman-mackey2013}, Matplotlib \citep{hunter2007}, NumPy \citep{vanderwalt2011}, PypeIt \citep{prochaska2020}, sbpy \citep{mommert2019}}

\bibliography{ms}

\begin{thebibliography}{}
\expandafter\ifx\csname natexlab\endcsname\relax\def\natexlab#1{#1}\fi
\providecommand{\url}[1]{\href{#1}{#1}}
\providecommand{\dodoi}[1]{doi:~\href{http://doi.org/#1}{\nolinkurl{#1}}}
\providecommand{\doeprint}[1]{\href{http://ascl.net/#1}{\nolinkurl{http://ascl.net/#1}}}
\providecommand{\doarXiv}[1]{\href{https://arxiv.org/abs/#1}{\nolinkurl{https://arxiv.org/abs/#1}}}

\bibitem[{A'Hearn {et~al.}(1995)A'Hearn, Millis, Schleicher, Osip, \&
  Birch}]{ahearn1995}
A'Hearn, M.~F., Millis, R.~C., Schleicher, D.~G., Osip, D.~J., \& Birch, P.~V.
  1995, Icar, 118, 223

\bibitem[{A'Hearn {et~al.}(1984)A'Hearn, Schleicher, Millis, Feldman, \&
  Thompson}]{ahearn1984}
A'Hearn, M.~F., Schleicher, D., Millis, R., Feldman, P., \& Thompson, D. 1984,
  AJ, 89, 579

\bibitem[{{Astropy Collaboration} {et~al.}(2013){Astropy Collaboration},
  Robitaille, Tollerud, Greenfield, Droettboom, Bray, Aldcroft, Davis,
  Ginsburg, Price-Whelan, Kerzendorf, {et~al.}}]{astropy2013}
{Astropy Collaboration}, Robitaille, T.~P., Tollerud, E.~J., {et~al.} 2013,
  A\&A, 558, A33

\bibitem[{Beech(1998)}]{beech1998}
Beech, M. 1998, MNRAS, 294, 259

\bibitem[{Benna {et~al.}(2015)Benna, Mahaffy, Grebowsky, Plane, Yelle, \&
  Jakosky}]{benna2015}
Benna, M., Mahaffy, P., Grebowsky, J., {et~al.} 2015, GeoRL, 42, 4670

\bibitem[{Boe {et~al.}(2019)Boe, Jedicke, Meech, Wiegert, Weryk, Chambers,
  Denneau, Kaiser, Kudritzki, Magnier, {et~al.}}]{boe2019}
Boe, B., Jedicke, R., Meech, K.~J., {et~al.} 2019, Icar, 333, 252

\bibitem[{Cabannes {et~al.}(1938)Cabannes, Dufay, \& Gauzit}]{cabannes1938}
Cabannes, J., Dufay, J., \& Gauzit, J. 1938, ApJ, 88, 164

\bibitem[{Carbary {et~al.}(2004)Carbary, Morrison, Romick, \&
  Yee}]{carbary2004}
Carbary, J., Morrison, D., Romick, G., \& Yee, J. 2004, ASR, 33, 1455

\bibitem[{Chamberlain(1956)}]{chamberlain1956}
Chamberlain, J.~W. 1956, JATP, 9, 73

\bibitem[{Chambers {et~al.}(2016)Chambers, Magnier, Metcalfe, Flewelling,
  Huber, Waters, Denneau, Draper, Farrow, Finkbeiner, {et~al.}}]{chambers2016}
Chambers, K., Magnier, E., Metcalfe, N., {et~al.} 2016, arXiv:1612.05560

\bibitem[{Chapman(1939)}]{chapman1939}
Chapman, S. 1939, ApJ, 90, 309

\bibitem[{Christou(2010)}]{christou2010}
Christou, A. 2010, MNRAS, 402, 2759

\bibitem[{Christou(2004)}]{christou2004}
Christou, A.~A. 2004, Icar, 168, 23

\bibitem[{Combi {et~al.}(2019)Combi, M{\"a}kinen, Bertaux, Qu{\'e}merais,
  Ferron, \& Coronel}]{combi2019}
Combi, M.~R., M{\"a}kinen, T., Bertaux, J.-L., {et~al.} 2019, ApJL, 884, L39

\bibitem[{Crismani {et~al.}(2017)Crismani, Schneider, Plane, Evans, Jain,
  Chaffin, Carrillo-Sanchez, Deighan, Yelle, Stewart, {et~al.}}]{crismani2017}
Crismani, M.~M., Schneider, N.~M., Plane, J.~M., {et~al.} 2017, NatGe, 10, 401

\bibitem[{Cui {et~al.}(2012)Cui, Zhao, Chu, Li, Li, Zhang, Su, Yao, Wang, Xing,
  {et~al.}}]{cui2012}
Cui, X.-Q., Zhao, Y.-H., Chu, Y.-Q., {et~al.} 2012, RAA, 12, 1197

\bibitem[{DiSanti {et~al.}(2016)DiSanti, Bonev, Gibb, Paganini, Villanueva,
  Mumma, Keane, Blake, Russo, Meech, {et~al.}}]{disanti2016}
DiSanti, M., Bonev, B., Gibb, E., {et~al.} 2016, ApJ, 820, 34

\bibitem[{Farnham {et~al.}(2000)Farnham, Schleicher, \& A'Hearn}]{farnham2000}
Farnham, T.~L., Schleicher, D.~G., \& A'Hearn, M.~F. 2000, Icar, 147, 180

\bibitem[{Farnham {et~al.}(2001)Farnham, Schleicher, Woodney, Birch, Eberhardy,
  \& Levy}]{farnham2001}
Farnham, T.~L., Schleicher, D.~G., Woodney, L.~M., {et~al.} 2001, Sci, 292,
  1348

\bibitem[{Farnocchia {et~al.}(2014)Farnocchia, Chesley, Chodas, Tricarico,
  Kelley, \& Farnham}]{farnocchia2014}
Farnocchia, D., Chesley, S.~R., Chodas, P.~W., {et~al.} 2014, ApJ, 790, 114

\bibitem[{Farnocchia {et~al.}(2016)Farnocchia, Chesley, Micheli, Delamere,
  Heyd, Tholen, Giorgini, Owen, \& Tamppari}]{farnocchia2016}
Farnocchia, D., Chesley, S.~R., Micheli, M., {et~al.} 2016, Icar, 266, 279

\bibitem[{Feldman {et~al.}(2018)Feldman, Weaver, A’Hearn, Combi, \&
  Russo}]{feldman2018}
Feldman, P.~D., Weaver, H.~A., A’Hearn, M.~F., Combi, M.~R., \& Russo, N.~D.
  2018, AJ, 155, 193

\bibitem[{Finson \& Probstein(1968)}]{finson1968}
Finson, M., \& Probstein, R. 1968, ApJ, 154, 327

\bibitem[{Foreman-Mackey {et~al.}(2013)Foreman-Mackey, Hogg, Lang, \&
  Goodman}]{foreman-mackey2013}
Foreman-Mackey, D., Hogg, D.~W., Lang, D., \& Goodman, J. 2013, PASP, 125, 306

\bibitem[{Fray {et~al.}(2005)Fray, B{\'e}nilan, Cottin, Gazeau, \&
  Crovisier}]{fray2005}
Fray, N., B{\'e}nilan, Y., Cottin, H., Gazeau, M.-C., \& Crovisier, J. 2005,
  P\&SS, 53, 1243

\bibitem[{Fulle(2004)}]{fulle2004}
Fulle, M. 2004, in Comets II, ed. M.~C. Festou, H.~U. Keller, \& H.~A. Weaver
  (Univ. of Arizona Press), 565--576

\bibitem[{{Gaia Collaboration} {et~al.}(2021){Gaia Collaboration}, Brown,
  Vallenari, Prusti, De~Bruijne, Babusiaux, Biermann, {et~al.}}]{gaia2021}
{Gaia Collaboration}, Brown, A.~G., Vallenari, A., {et~al.} 2021, A\&A

\bibitem[{H{\"a}nni {et~al.}(2020)H{\"a}nni, Altwegg, Pestoni, Rubin,
  Schroeder, Schuhmann, \& Wampfler}]{hanni2020}
H{\"a}nni, N., Altwegg, K., Pestoni, B., {et~al.} 2020, MNRAS, 498, 2239

\bibitem[{Hansell {et~al.}(1995)Hansell, Wells, \& Hunten}]{hansell1995}
Hansell, S., Wells, W., \& Hunten, D. 1995, Icar, 117, 345

\bibitem[{Haser(1957)}]{haser1957}
Haser, L. 1957, BSRSL, 43, 740

\bibitem[{Hui {et~al.}(2017)Hui, Jewitt, \& Clark}]{hui2017}
Hui, M.-T., Jewitt, D., \& Clark, D. 2017, AJ, 155, 25

\bibitem[{Hunter(2007)}]{hunter2007}
Hunter, J.~D. 2007, CSE, 9, 90

\bibitem[{Imamura {et~al.}(2017)Imamura, Ando, Tellmann, P{\"a}tzold,
  H{\"a}usler, Yamazaki, Sato, Noguchi, Futaana, Oschlisniok,
  {et~al.}}]{imamura2017}
Imamura, T., Ando, H., Tellmann, S., {et~al.} 2017, EP\&S, 69, 1

\bibitem[{Jenniskens {et~al.}(2021)Jenniskens, Lauretta, Towner, Heathcote,
  Jehin, Hanke, Cooper, Baggaley, Howell, Johannink, {et~al.}}]{jenniskens2021}
Jenniskens, P., Lauretta, D.~S., Towner, M.~C., {et~al.} 2021, Icar, 365,
  114469

\bibitem[{Jewitt(2015)}]{jewitt2015}
Jewitt, D. 2015, AJ, 150, 201

\bibitem[{Jewitt {et~al.}(2019)Jewitt, Agarwal, Hui, Li, Mutchler, \&
  Weaver}]{jewitt2019}
Jewitt, D., Agarwal, J., Hui, M.-T., {et~al.} 2019, AJ, 157, 65

\bibitem[{Jewitt {et~al.}(2021)Jewitt, Kim, Mutchler, Agarwal, Li, \&
  Weaver}]{jewitt2021}
Jewitt, D., Kim, Y., Mutchler, M., {et~al.} 2021, AJ, 161, 188

\bibitem[{Jewitt {et~al.}(1996)Jewitt, Luu, \& Chen}]{jewitt1996}
Jewitt, D., Luu, J., \& Chen, J. 1996, AJ, 112, 1225

\bibitem[{Jewitt \& Matthews(1999)}]{jewitt1999}
Jewitt, D., \& Matthews, H. 1999, AJ, 117, 1056

\bibitem[{Jones {et~al.}(2018)Jones, Knight, Battams, Boice, Brown, Giordano,
  Raymond, Snodgrass, Steckloff, Weissman, {et~al.}}]{jones2018}
Jones, G.~H., Knight, M.~M., Battams, K., {et~al.} 2018, SSRv, 214, 20

\bibitem[{Kelley {et~al.}(2014)Kelley, Farnham, Bodewits, Tricarico, \&
  Farnocchia}]{kelley2014}
Kelley, M.~S., Farnham, T.~L., Bodewits, D., Tricarico, P., \& Farnocchia, D.
  2014, ApJL, 792, L16

\bibitem[{Kizner(1961)}]{kizner1961}
Kizner, W. 1961, P\&SS, 7, 125

\bibitem[{Knight \& Schleicher(2014)}]{knight2014}
Knight, M.~M., \& Schleicher, D.~G. 2014, AJ, 149, 19

\bibitem[{Kolokolova {et~al.}(2018)Kolokolova, Nagdimunov, \&
  Mackowski}]{kolokolova2018}
Kolokolova, L., Nagdimunov, L., \& Mackowski, D. 2018, JQSRT, 204, 138

\bibitem[{Krasnopol'sky(1983)}]{krasnopolsky1983}
Krasnopol'sky, V. 1983, in Venus, ed. D.~M. Hunten, L.~Colin, \& T.~M. Donahue
  (Univ. of Arizona Press), 459--483

\bibitem[{Kresak(1993)}]{kresak1993}
Kresak, L. 1993, A\&A, 279, 646

\bibitem[{Kr{\'o}likowska \& Dybczy{\'n}ski(2017)}]{krolikowska2017}
Kr{\'o}likowska, M., \& Dybczy{\'n}ski, P.~A. 2017, MNRAS, 472, 4634

\bibitem[{Lasue {et~al.}(2009)Lasue, Levasseur-Regourd, Hadamcik, \&
  Alcouffe}]{lasue2009}
Lasue, J., Levasseur-Regourd, A.~C., Hadamcik, E., \& Alcouffe, G. 2009, Icar,
  199, 129

\bibitem[{Leonard {et~al.}(2021{\natexlab{a}})Leonard, Aschi, Pettarin,
  Groeller, Rankin, Gray, Shelly, Christensen, Ries, Roman,
  {et~al.}}]{leonard2021a}
Leonard, G., Aschi, S., Pettarin, E., {et~al.} 2021{\natexlab{a}}, MPEC,
  2021-A99

\bibitem[{Leonard {et~al.}(2021{\natexlab{b}})Leonard, Aschi, Pettarin,
  Groeller, Rankin, Gray, Shelly, Christensen, Ries, Roman,
  {et~al.}}]{leonard2021b}
---. 2021{\natexlab{b}}, CBET, 4907

\bibitem[{Li \& Jewitt(2015)}]{li2015}
Li, J., \& Jewitt, D. 2015, AJ, 149, 133

\bibitem[{Li {et~al.}(2014)Li, Samarasinha, Kelley, Farnham, A'Hearn, Mutchler,
  Lisse, \& Delamere}]{li2014}
Li, J.-Y., Samarasinha, N.~H., Kelley, M.~S., {et~al.} 2014, ApJL, 797, L8

\bibitem[{Li {et~al.}(2016)Li, Samarasinha, Kelley, Farnham, Bodewits, Lisse,
  Mutchler, A’Hearn, \& Delamere}]{li2016}
---. 2016, ApJL, 817, L23

\bibitem[{Lyytinen \& Jenniskens(2003)}]{lyytinen2003}
Lyytinen, E., \& Jenniskens, P. 2003, Icar, 162, 443

\bibitem[{Marsden(1988)}]{marsden1988}
Marsden, B. 1988, BAAS, 20, 898

\bibitem[{Massey {et~al.}(2013)Massey, Dunham, Bida, Collins, Hall, Hunter,
  Lauman, Levine, Neugent, Nye, {et~al.}}]{massey2013}
Massey, P., Dunham, E., Bida, T., {et~al.} 2013, AAS, 221, 345.02

\bibitem[{McAuliffe \& Christou(2006)}]{mcauliffe2006}
McAuliffe, J.~P., \& Christou, A.~A. 2006, Icar, 180, 8

\bibitem[{Meech \& Svoren(2004)}]{meech2004}
Meech, K., \& Svoren, J. 2004, in Comets II, ed. M.~C. Festou, H.~U. Keller, \&
  H.~A. Weaver (Univ. of Arizona Press), 317--336

\bibitem[{Meech {et~al.}(2017)Meech, Kleyna, Hainaut, Micheli, Bauer, Denneau,
  Keane, Stephens, Jedicke, Wainscoat, {et~al.}}]{meech2017}
Meech, K.~J., Kleyna, J.~T., Hainaut, O., {et~al.} 2017, ApJL, 849, L8

\bibitem[{Meftah {et~al.}(2018)Meftah, Dam{\'e}, Bols{\'e}e, Hauchecorne,
  Pereira, Sluse, Cessateur, Irbah, Bureau, Weber, {et~al.}}]{meftah2018}
Meftah, M., Dam{\'e}, L., Bols{\'e}e, D., {et~al.} 2018, A\&A, 611, A1

\bibitem[{Miles(2013)}]{miles2013}
Miles, R. 2013, JBAA, 123, 363

\bibitem[{Mommert {et~al.}(2019)Mommert, Kelley, de~Val-Borro, Li, Guzman,
  vDurech, Granvik, Grundy, Moskovitz, Penttil{\"a}, {et~al.}}]{mommert2019}
Mommert, M., Kelley, M., de~Val-Borro, M., {et~al.} 2019, JOSS, 4

\bibitem[{Moorhead {et~al.}(2014)Moorhead, Wiegert, \& Cooke}]{moorhead2014}
Moorhead, A.~V., Wiegert, P.~A., \& Cooke, W.~J. 2014, Icar, 231, 13

\bibitem[{Nakamura {et~al.}(2011)Nakamura, Imamura, Ishii, Abe, Satoh, Suzuki,
  Ueno, Yamazaki, Iwagami, Watanabe, {et~al.}}]{nakamura2011}
Nakamura, M., Imamura, T., Ishii, N., {et~al.} 2011, EP\&S, 63, 443

\bibitem[{Ohtsuka(1991)}]{ohtsuka1991}
Ohtsuka, K. 1991, in Origin and Evolution of Interplanetary Dust, ed. A.~C.
  Levasseur-Regourd \& H.~Hasegawa (Springer), 315--318

\bibitem[{Oke \& Gunn(1982)}]{oke1982}
Oke, J., \& Gunn, J. 1982, PASP, 94, 586

\bibitem[{Opitom {et~al.}(2016)Opitom, Guilbert-Lepoutre, Jehin, Manfroid,
  Hutsem{\'e}kers, Gillon, Magain, Roberts-Borsani, \& Witasse}]{opitom2016}
Opitom, C., Guilbert-Lepoutre, A., Jehin, E., {et~al.} 2016, A\&A, 589, A8

\bibitem[{P{\"a}tzold {et~al.}(2009)P{\"a}tzold, Tellmann, H{\"a}usler, Bird,
  Tyler, Christou, \& Withers}]{patzold2009}
P{\"a}tzold, M., Tellmann, S., H{\"a}usler, B., {et~al.} 2009, GRL, 36

\bibitem[{Prochaska {et~al.}(2020)Prochaska, Hennawi, Westfall, Cooke, Wang,
  Hsyu, Davies, Farina, \& Pelliccia}]{prochaska2020}
Prochaska, J.~X., Hennawi, J.~F., Westfall, K.~B., {et~al.} 2020, JOSS, 5, 2308

\bibitem[{Rauer {et~al.}(1997)Rauer, Arpigny, Boehnhardt, Colas, Crovisier,
  Jorda, K{\"u}ppers, Manfroid, Rembor, \& Thomas}]{rauer1997}
Rauer, H., Arpigny, C., Boehnhardt, H., {et~al.} 1997, Sci, 275, 1909

\bibitem[{Restano {et~al.}(2015)Restano, Plaut, Campbell, Gim, Nunes,
  Bernardini, Egan, Seu, \& Phillips}]{restano2015}
Restano, M., Plaut, J.~J., Campbell, B.~A., {et~al.} 2015, GRL, 42, 4663

\bibitem[{Rousselot(2008)}]{rousselot2008}
Rousselot, P. 2008, A\&A, 480, 543

\bibitem[{Russo {et~al.}(2008)Russo, Vervack~Jr, Weaver, Montgomery, Deshpande,
  Fern{\'a}ndez, \& Martin}]{russo2008}
Russo, N.~D., Vervack~Jr, R., Weaver, H., {et~al.} 2008, ApJ, 680, 793

\bibitem[{S{\'a}nchez-Cano {et~al.}(2020)S{\'a}nchez-Cano, Lester, Witasse,
  Morgan, Opgenoorth, Andrews, Blelly, Cowley, Kopf, Leblanc,
  {et~al.}}]{sanchezcano2020}
S{\'a}nchez-Cano, B., Lester, M., Witasse, O., {et~al.} 2020, JGRA, 125,
  e2019JA027344

\bibitem[{Schleicher \& Bair(2011)}]{schleicher2011}
Schleicher, D.~G., \& Bair, A.~N. 2011, AJ, 141, 177

\bibitem[{Schleicher {et~al.}(2002)Schleicher, Woodney, \&
  Birch}]{schleicher2002}
Schleicher, D.~G., Woodney, L.~M., \& Birch, P.~V. 2002, in Cometary Science
  after Hale-Bopp (Springer), 401--403

\bibitem[{Schneider {et~al.}(2015)Schneider, Deighan, Stewart, McClintock,
  Jain, Chaffin, Stiepen, Crismani, Plane, Carrillo-S{\'a}nchez,
  {et~al.}}]{schneider2015}
Schneider, N.~M., Deighan, J., Stewart, A., {et~al.} 2015, GeoRL, 42, 4755

\bibitem[{Sekanina \& Chodas(2004)}]{sekanina2004}
Sekanina, Z., \& Chodas, P.~W. 2004, ApJ, 607, 620

\bibitem[{Sekanina \& Kracht(2014)}]{sekanina2014}
Sekanina, Z., \& Kracht, R. 2014, arXiv:1404.5968

\bibitem[{Sekanina \& Kracht(2016)}]{sekanina2016}
---. 2016, ApJ, 823, 2

\bibitem[{Shrben{\`y} \& Spurn{\`y}(2009)}]{shrbeny2009}
Shrben{\`y}, L., \& Spurn{\`y}, P. 2009, A\&A, 506, 1445

\bibitem[{Skrutskie {et~al.}(2006)Skrutskie, Cutri, Stiening, Weinberg,
  Schneider, Carpenter, Beichman, Capps, Chester, Elias,
  {et~al.}}]{skrutskie2006}
Skrutskie, M., Cutri, R., Stiening, R., {et~al.} 2006, AJ, 131, 1163

\bibitem[{Slanger {et~al.}(2001)Slanger, Cosby, Huestis, \& Bida}]{slanger2001}
Slanger, T., Cosby, P., Huestis, D., \& Bida, T. 2001, Sci, 291, 463

\bibitem[{Sykes \& Walker(1992)}]{sykes1992}
Sykes, M.~V., \& Walker, R.~G. 1992, Icar, 95, 180

\bibitem[{Takahashi {et~al.}(2018)Takahashi, Sato, Imai, Lorenz, Yair, Aplin,
  Fischer, Nakamura, Ishii, Abe, {et~al.}}]{takahashi2018}
Takahashi, Y., Sato, M., Imai, M., {et~al.} 2018, EP\&S, 70, 1

\bibitem[{Tricarico {et~al.}(2014)Tricarico, Samarasinha, Sykes, Li, Farnham,
  Kelley, Farnocchia, Stevenson, Bauer, \& Lock}]{tricarico2014}
Tricarico, P., Samarasinha, N.~H., Sykes, M.~V., {et~al.} 2014, ApJL, 787, L35

\bibitem[{Trigo-Rodr{\'\i}guez {et~al.}(2008)Trigo-Rodr{\'\i}guez,
  Garc{\'\i}a-Melendo, Davidsson, S{\'a}nchez, Rodr{\'\i}guez, Lacruz,
  de~Los~Reyes, \& Pastor}]{trigo2008}
Trigo-Rodr{\'\i}guez, J., Garc{\'\i}a-Melendo, E., Davidsson, B., {et~al.}
  2008, A\&A, 485, 599

\bibitem[{Valsecchi {et~al.}(2003)Valsecchi, Milani, Gronchi, \&
  Chesley}]{valsecchi2003}
Valsecchi, G.~B., Milani, A., Gronchi, G.~F., \& Chesley, S.~R. 2003, A\&A,
  408, 1179

\bibitem[{Van Der~Walt {et~al.}(2011)Van Der~Walt, Colbert, \&
  Varoquaux}]{vanderwalt2011}
Van Der~Walt, S., Colbert, S.~C., \& Varoquaux, G. 2011, CSE, 13, 22

\bibitem[{Vaubaillon {et~al.}(2014)Vaubaillon, Maquet, \&
  Soja}]{vaubaillon2014}
Vaubaillon, J., Maquet, L., \& Soja, R. 2014, MNRAS, 439, 3294

\bibitem[{Wilson {et~al.}(2003)Wilson, Eikenberry, Henderson, Hayward, Carson,
  Pirger, Barry, Brandl, Houck, Fitzgerald, {et~al.}}]{wilson2003}
Wilson, J.~C., Eikenberry, S.~S., Henderson, C.~P., {et~al.} 2003, Proc. SPIE,
  4841, 451

\bibitem[{Withers {et~al.}(2013)Withers, Christou, \& Vaubaillon}]{withers2013}
Withers, P., Christou, A., \& Vaubaillon, J. 2013, ASR, 52, 1207

\bibitem[{Yamazaki {et~al.}(2018)Yamazaki, Yamada, Lee, Watanabe, Horinouchi,
  Murakami, Kouyama, Ogohara, Imamura, Sato, {et~al.}}]{yamazaki2018}
Yamazaki, A., Yamada, M., Lee, Y.~J., {et~al.} 2018, EP\&S, 70, 1

\bibitem[{Yang {et~al.}(2021)Yang, Jewitt, Zhao, Jiang, Ye, \& Chen}]{yang2021}
Yang, B., Jewitt, D., Zhao, Y., {et~al.} 2021, arXiv:2105.10986

\bibitem[{Ye {et~al.}(2021)Ye, Knight, Kelley, Moskovitz, Gustafsson, \&
  Schleicher}]{ye2021}
Ye, Q., Knight, M.~M., Kelley, M.~S., {et~al.} 2021, PSJ, 2, 23

\bibitem[{Ye {et~al.}(2015)Ye, Brown, Bell, Gao, Ma{\v{s}}ek, \& Hui}]{ye2015}
Ye, Q.-Z., Brown, P.~G., Bell, C., {et~al.} 2015, ApJ, 814, 79

\bibitem[{Ye \& Hui(2014)}]{ye2014}
Ye, Q.-Z., \& Hui, M.-T. 2014, ApJ, 787, 115

\end{thebibliography}

\end{CJK*}
\end{document}